\documentclass[12pt, onecolumn]{aa}
\def\@@ptsize{16pt}

\usepackage{natbib}
\usepackage{graphicx}
\usepackage{amssymb}
\usepackage{amsmath}
\usepackage{amsfonts}
\usepackage{url}
\usepackage{hyperref}
\usepackage{mathptmx}
\usepackage{alltt}
\usepackage{listings}
\usepackage{lineno}
\usepackage{txfonts}
\RequirePackage{array}
\usepackage{makeidx}
\usepackage{multirow}
\usepackage[table]{xcolor}
\usepackage{booktabs,xspace}
\usepackage[T1]{fontenc}
\usepackage[font=small,labelfont=bf,tableposition=top]{caption}

\DeclareCaptionLabelFormat{andtable}{#1~#2  \&  \tablename~\thetable}

\renewcommand{\vec}[1]{\ensuremath{\mathchoice{\mbox{\boldmath$\displaystyle#1$}}
		{\mbox{\boldmath$\textstyle#1$}}
		{\mbox{\boldmath$\scriptstyle#1$}}
		{\mbox{\boldmath$\scriptscriptstyle#1$}}}}

\def\apjl{ApJ}%
\def\apjs{ApJS}%
\def\ao{Appl.~Opt.}%
\def\aap{A\&A}%
\begin{document}
	\pagestyle{plain}
	\pagenumbering{arabic}
	

\begin{center}
	{\LARGE{\bf
			{Proposal title: GaiaNIR \\ Combining optical and Near-Infra-Red (NIR) capabilities with Time-Delay-Integration (TDI) sensors for a future Gaia-like mission.}
	}}
\end{center}

\bigskip
	
\begin{center}
	{\Large{\bf PI: 
		{Dr. David Hobbs, \\ Lund Observatory, Box 43, SE-221 00 Lund, Sweden. \\ Email: david@astro.lu.se.~~~Tel.: +46-46-22\,21573}
	}}
\end{center}

\begin{tabular}{ll}
	Core team members: & The following minimum team is needed to initiate the project.\\ 
	D.\ Hobbs          & Lund Observatory, Sweden. \\
	A.\ Brown          & Leiden Observatory, Holland.\\
	A.\ Mora           & Aurora Technology B.V., Spain. \\
	C.\ Crowley        & HE Space Operations B.V., Spain. \\
	N.\ Hambly         & University of Edinburgh, UK. \\
	J.\ Portell        & Institut de Ci\`encies del Cosmos, ICCUB-IEEC, Spain. \\
	C.\ Fabricius      & Institut de Ci\`encies del Cosmos, ICCUB-IEEC, Spain. \\
	M.\ Davidson       & University of Edinburgh, UK. \\ \\

	Proposal writers:            & See Appendix \ref{Sec:AppA}.\\ \\
	Other supporting scientists: & See Appendix \ref{Sec:AppB} and Appendix \ref{Sec:AppC}.\\ \\

	Senior science advisors:  &\\
	E.\ H{\o}g                & Copenhagen University (Retired), Denmark. \\
	L.\ Lindegren             & Lund Observatory, Sweden. \\
	C.\ Jordi                 & Institut de Ci\`encies del Cosmos, ICCUB-IEEC, Spain. \\
	S.\ Klioner               & Lohrmann Observatory, Germany.\\
	F.\ Mignard               & Observatoire de la C\^{o}te d'Azur, France.\\ 
\end{tabular}

\bigskip

\begin{center}
\begin{figure}[tbh]
	\includegraphics[scale=0.314,angle=90]{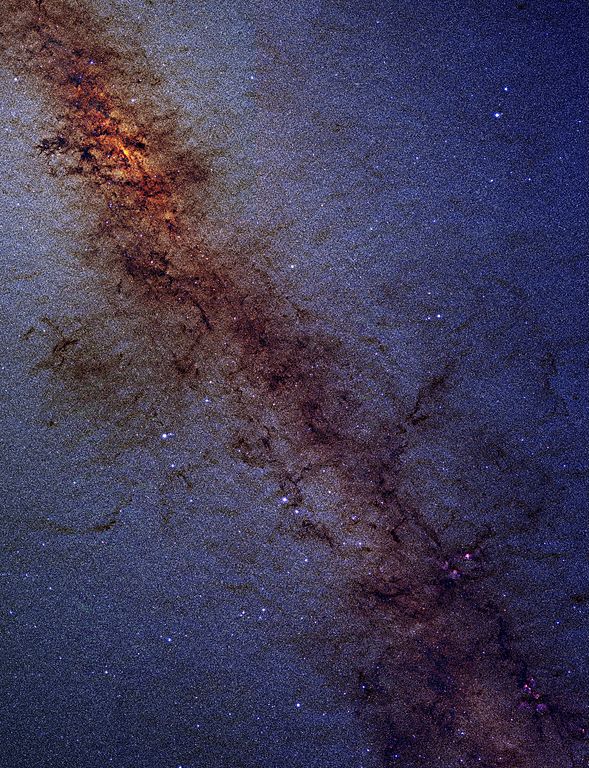}
	\includegraphics[scale=0.823]{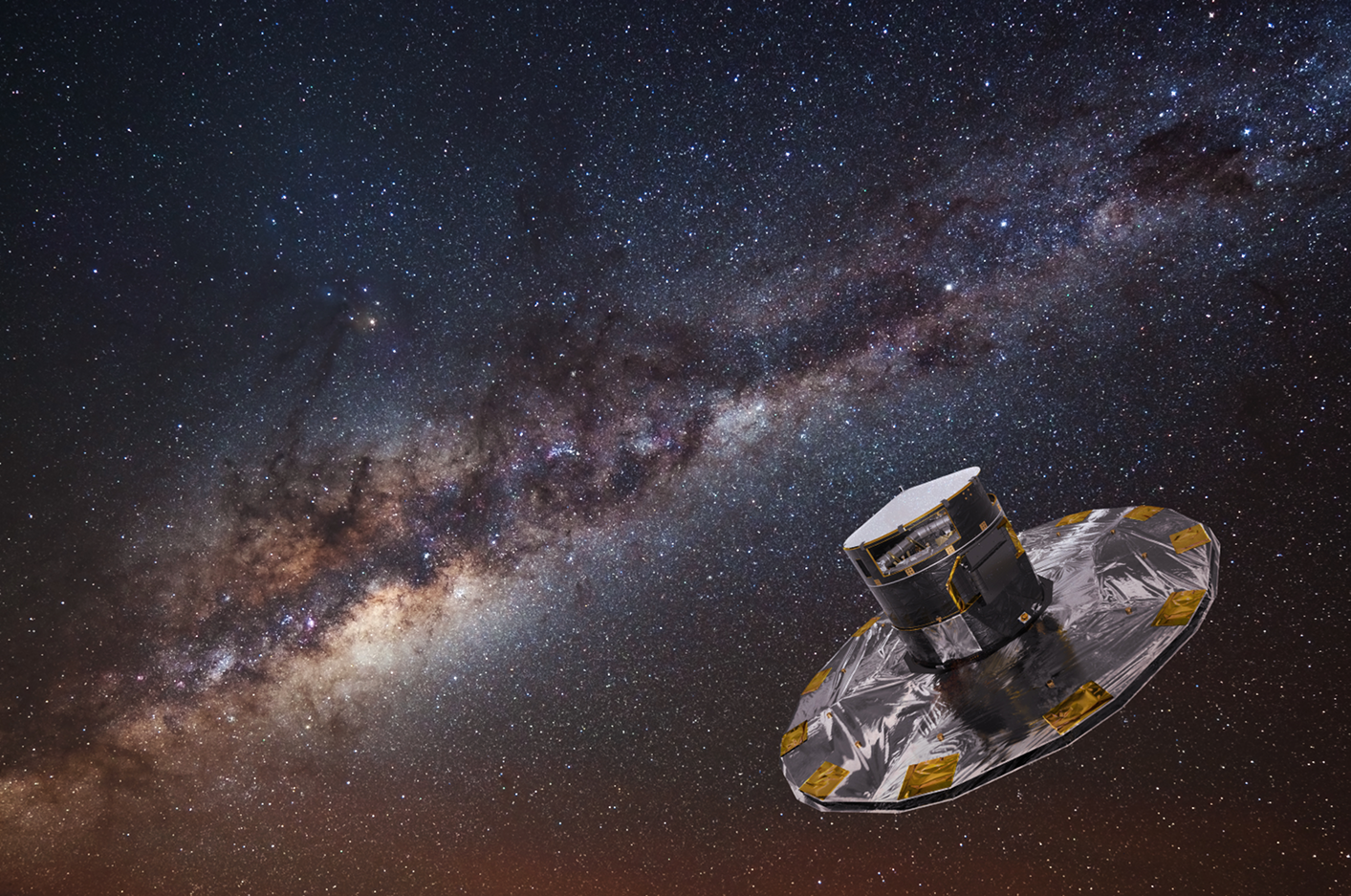}
	\caption{\em{Left is an IR image from the Two Micron All-Sky Survey (image G. Kopan, R. Hurt) while on the right an artist's concept of the Gaia mission superimposed on an optical image, (Image ESA). Images not to scale.
}}
\end{figure}
\end{center}

\newpage
\section{Executive summary}\label{summary}
ESA recently called for new ``Science Ideas'' to be investigated in terms of feasibility and 
technological developments -- for technologies not yet sufficiently mature. These ideas may in 
the future become candidates for M or L class missions within the ESA Science Program.

With the launch of Gaia in December 2013, Europe entered a new era of space astrometry 
following in the footsteps of the very successful Hipparcos mission from the early 1990s. 
Gaia is the successor to Hipparcos, both of which operated in optical wavelengths, and Gaia is two 
orders of magnitude more accurate in the five astrometric parameters and is surveying four orders 
of magnitude more stars in a vast volume of the Milky Way. The combination of the Hipparcos/Tycho-2 catalogues 
with the first early Gaia data release will give improved proper motions over a long $\sim$25 year baseline. 
The final Gaia  solution will also establish a new optical reference frame by means of quasars, by linking the 
optical counterparts of radio (VLBI) sources defining the orientation of the reference frame, and 
by using the zero proper motion of quasars to determine a non-rotating frame \citep{2012A&A...538A..78L}.

A weakness of Gaia is that it only operates at optical wavelengths.  However, much of the Galactic centre and 
the spiral arm regions, important for certain studies, are obscured by interstellar extinction and this makes 
it difficult for Gaia to deeply probe. Traditionally, this 
problem is overcome by switching to the infra-red but this was not possible with Gaia's CCDs. Additionally,
to scan the entire sky and make global absolute parallax measurements the spacecraft must have a constant 
rotation and this requires that the CCDs operate in TDI mode, increasing their complexity. Two main 
scientific motivations for a new Gaia-like mission are:

\begin{enumerate}
\item An astrometric Gaia-like successor mission has great and unique scientific interest. Proper motions with fourteen times 
smaller errors than from Gaia alone will be obtained when positions from two epochs at 20 year intervals are combined. 
This opens a number of science cases which cannot be tackled in any other way. At the same time the parallaxes, especially 
of binaries, will be much improved when astrometric data from two missions are combined. A new mission would also allow the 
slowly degrading accuracy of the Gaia optical reference frame, which will be the basis for modern astronomical
measurements, to be maintained.

\item Near-Infra-Red (NIR) astrometry (and simultaneous NIR photometry) is interesting for penetrating obscured regions and 
for observing intrinsically red objects if it can be implemented with sufficient accuracy. We know there are technological 
obstacles for NIR astrometry and photometry with TDI, e.g. large read out noise, and this is the reason to have an early 
technological study and possible detector development.
\end{enumerate}

This proposal has evolved in a number of people's minds over recent years, particularly within the Gaia community. Notably, Erik H{\o}g 
has collected many detailed ideas in a series of technical notes outlining various science cases and instrument designs for a Gaia-like 
successor mission. In 2013 both Erik and Anthony Brown wrote white papers for the definition of the L2 and L3 missions in the ESA Science 
Programme entitled ``Astrometry for Dynamics'' and ``Space-Time Structure Explorer Sub-Microarcsecond Astrometry for the 2030s'' 
respectively. From these white papers, technical notes and other sources three core ideas for future astrometry have emerged:
\begin{enumerate}
\item Nano-arcsecond astrometry (Pointed mission)
\item Global optical astrometry (Gaia2)
\item Global optical and near-infra-red astrometry (GaiaNIR)
\end{enumerate}
In July 2015 a meeting was organized in Cambridge entitled ``Next Steps Towards Future Space Astrometry 
Missions''\footnote{The web page for details and references is available at this \href{http://great.ast.cam.ac.uk/Greatwiki/GaiaScienceMeetings/FutureAstrometryJul15\#Next_Steps_Towards_Future_Space_Astrometry_Missions}{link}.}
in which possible future missions were discussed. Based on this background the current proposal collects together the science cases and 
instrument designs relevant to pursue the third option listed above, namely the development of NIR detectors with TDI mode for future 
global space astrometry missions.

The purpose of this proposal is to initiate an industrial scientific study to determine how best to expand the 
sensitivity of a Gaia-like mission into the NIR to give a potential wavelength range of 400--2000~nm. HgCdTe (also known as MCT) 
materials seem to be currently the most promising for NIR sensors. Hybrid HgCdTe-CMOS or InGaAs sensors are suggested as possibilities but 
their readout noise, achievable astrometric accuracy, cooling strategy, etc. need to be assessed. Another key question is whether the 
focal plane consist of just one type of detector or should it be combined with conventional optical CCD detectors. Early 
studies\footnote{See this \href{https://dl.dropboxusercontent.com/u/49240691/GaiaNIRCorr.pdf}{link}.} indicated that the main scientific
goals could still be achieved using a NIR capability in a limited sensitivity range from 400 to 1600 nm or even only from 900 to 1600 nm
if also combined with optical CCDs.

The science case is to build on the Gaia results of all-sky absolute astrometry for more than one billion stars. The Gaia successor 
should provide astrometry with equal or better accuracy for the same stars but will also observe many new stars in obscured regions. 
A common solution of the data of the two missions will give greatly improved proper motions but also improve the parallaxes 
by doubling the number of observations (and by having the better proper motions) for common stars. Long term maintenance of a dense 
and very accurate celestial reference frame with a new mission is necessary for future precise astronomical observations. 
Pushing into the NIR opens up a new wavelength range which allows us to probe the dusty obscured regions of the Galaxy with 
high-precision astrometry and broad-band high-resolution photometry. 

In the long run the same kind of technology could pave the way for a panchromatic astrometric and photometric global surveyor, 
which could be another mission on its own, covering from the ultraviolet (UV) to the NIR or even the mid-IR with mmag absolute 
photometric accuracy, and an all-sky nature that is impossible to achieve on-ground. 

\newpage
\section{Science case}\label{sciencecase}

The current Gaia mission will provide global astrometry (20--25~$\mu$as at G=15), absolute parallaxes and proper motions, to 
unprecedented accuracies, with the addition of all-sky homogeneous multi-colour photometry and spectroscopy. 
These unique capabilities go well beyond and are complementary to the science cases being addressed by ground based 
surveys (such as RAVE, SDSS, Pan-Starrs, APOGEE, LSST, etc). A space-based mission avoids the limitations caused by the turbulent
atmosphere and the use of Earth rotation parameters and models of nutation and precession. All-sky space-based astrometry
leads to a global solution and provides a rigid sphere for a celestial reference frame that cannot be accurately obtained 
with any other method.

After the publication of the final Gaia catalogue the positions of stars will be accurately known at the chosen reference 
epoch (currently ~2017). However, this accurate positional information and the accuracy of the link to the VLBI reference 
frame will slowly degrade due to the small uncertainties in the proper motions of the stars. Hence, it is very desirable 
to repeat the measurements of Gaia after about 20 years to maintain the positional accuracy of the stars and the optical 
reference frame. Additionally, having two 5 or 10-year Gaia-like missions separated by 20 years would give 14 times 
better proper motions for more than one billion stars and also improved parallax determinations with new observations. 
An obvious technological improvement to the current Gaia mission is to also go into the Near-Infra-Red (NIR) allowing the 
new mission to probe through the Galactic dust to observe the structure and kinematics of the star forming regions in the disk, 
the spiral arms and the bulge region to give model independent distances and proper motions in these obscured parts of the sky.  

If we are to propose a new Gaia-like mission, with all-sky absolute optical and NIR astrometry, 
for launch around 2035 we can summarize a number of clear improvements to the science case that can be envisioned.
\begin{itemize}
\item Probing of the dusty Galactic centre and spiral arms to make the first microarcsecond astrometric measurements of these 
objects in the NIR and to connect them with known objects in a broader (optical) wavelength range and at a new epoch.  
\item A new mission could be combined with the older Gaia/Hipparcos/Tycho-2 catalogues to give a much 
longer baseline, with very accurate proper motions (a factor of 14 better in the two components).
Dynamical studies in the outer Galactic halo, where Gaia's proper motion accuracies are insufficient, would be greatly enhanced,
giving insight into dark matter. Kinematics of the local group of galaxies would also come within reach. 
\item Access to a deep survey of low-mass stars, brown dwarfs, and free floating planets over a large volume around the Sun. 
\item The longer baseline also allows longer periods (up to 40 years) to be probed for both exoplanets and binary stars which 
would much improve exoplanet detection accuracy and the characterization of the Galactic binary population.
\item Maintenance of a dense and very accurate celestial reference frame and its extension to the NIR to support, for example, 
future instruments like MICADO being developed for the E-ELT. To provide clear detection of the quasars solely from zero 
linear\footnote{Quasars will have some apparent proper motions due to time-dependent source structure. These proper motions are 
probably not linear and this can be used to distinguish them. Some nearby quasars will have normal proper motions
so spectra may still be needed to identify them.} proper motion and parallax, reducing the need for spectra.
\item Stars -- determination of astrophysical parameters for all stars observed, enhanced with new observations in the NIR.
\item The very accurate astrometry would greatly enhance our quantitative knowledge of Solar System objects.
\item To explore proper motion patterns, real time cosmology and fundamental physics.
\end{itemize}
				
The accuracy of the mission should be at least that of Gaia using tried and trusted instrumentation, 
techniques, and lessons learned from Gaia to unveil a wealth of new and more accurate information about our Galaxy.
To achieve these goals we need to explore the feasibility and technological developments needed to manufacture 
space qualified optical and NIR (400--2000~nm) TDI sensors with passive cooling. To maintain Europe's leadership in 
space astrometry it is highly desirable to develop such detector technology within Europe.
The most promising NIR sensors today seem to be hybrid HgCdTe-CMOS multiplexing sensors which can also support TDI 
mode. Using such detectors with extinction of A${\rm _V}=10$ magnitudes the same astrometric precision could 
be obtained for solar like stars that are 3--4 magnitudes fainter. A more limited sensitivity from 900 to 1600 nm in 
a part of the field of view may be considered if CCDs are also used for the astrometry and photometry. 
Alternatively, InGaAs hybrid sensors could be considered in view of the significantly higher readout noise in 
CMOS sensors which could limit their sensitivity to faint stars.

\subsection{Probing the dusty Galactic centre and spiral arms}\label{galacticcentre}

\subsubsection{The bulge/bar-disk interface and radial migration}\label{diskinterface}
Since Gaia was proposed, it has become clear that the evolution of the Milky Way is far more complicated than had been 
realized. Since stars  are radially migrated by bars \citep{1994ApJ...430L.105F} or spirals \citep{2002MNRAS.336..785S}, 
they can be found far from their birthplaces. This has important consequences for the chemical evolution of the Galaxy
\citep{2008ApJ...684L..79R, 2009MNRAS.399.1145S, 2010ApJ...722..112M}.
This radial migration means that we cannot understand components of the Galaxy in isolation. Stars born in the inner Galaxy 
can be found in the solar neighbourhood and vice versa. We cannot understand the history of the solar neighbourhood without 
understanding the history of the whole Galaxy.

An important element of radial migration happens at the interface between the bar and the spiral arms, which occurs in the 
inner few kpc of the Galaxy. In this region, extinction is a serious problem, and studies at visible wavelengths, such as 
Gaia, are unable to see through the dust to observe any but the intrinsically brightest and/or nearest stars.  An NIR 
astrometric mission would allow us to probe this region and determine proper motions and parallaxes of the stars 
there. This is the only way that we will be able to get a full 3D picture of the dynamics of this vitally important region 
of the Galaxy \citep[see][for a review]{2016arXiv160207702B}. 

\subsubsection{The bulge/bar}\label{bulgebar}
The centre of the Galaxy is dominated by a stellar bar, which has created a peanut-shaped pseudo-bulge \citep{2010ApJ...721L..28N}. There are also 
claims that the bar extends as far as 5kpc from the Galactic centre in the plane of the Galaxy \citep{2015MNRAS.450.4050W}. An NIR mission would 
provide proper motions and parallaxes for stars in this region and allow us to disentangle the dynamics of this complicated region. It would 
determine once and for all the structure of the inner Milky Way.

Cosmological models find that the density profiles of dark matter haloes peak at their centre with $\rho\propto r^{-1}$. However, this is 
in simulations assuming that dark matter is dynamically cold, and without any of the complicated physics associated with the baryonic components 
of galaxies. It has been shown that feedback from star formation can affect the dark matter density profile \citep{2012MNRAS.422.1231G}, and 
the effect of the bar on the dark matter halo is also the subject of debate \citep{2005MNRAS.363..991H}.

The Milky Way presents our best opportunity to study the dark matter content near the centre of an $L^\ast$ galaxy \citep{2005ApJ...627L..89C}, 
which may teach us about the nature of dark matter and/or the processes associated with galaxy formation and evolution on these scales. 
Without an astrometric mission that operates in the NIR, proper motions for these stars will never be determined, and the unique opportunity 
we have to study all-sky and the 6D phase space distribution of stars in the centre of an $L^\ast$ galaxy will be squandered.
However many of these objects will not have been observed by Gaia so that the GaiaNIR measurements will give only single epoch 
data. The parallaxes at the Galactic centre distances will thus not be very accurate for many new single objects. However, double epoch observations 
for bright stars will give a smaller sample of very accurate parallaxes and proper motions. Most ground based 
IR surveys, for example GRAVITY \citep{2011Msngr.143...16E}, will cover very small patches on the sky compared to the very unique all-sky 
astrometry that a Gaia-like mission can offer. While GRAVITY will be crucial to investigate better the orbits of stars around the super 
massive black hole at the Galactic centre, GaiaNIR will constrain the very detailed dynamical and orbital structure of many more stars and 
at greater distances around the black hole (i.e. at larger scales).

Similarly, the small-JASMINE/JASMINE missions currently being discussed in Japan\footnote{See this \href{http://www.scholarpedia.org/article/JASMINE\#Small-JASMINE}{link}.} 
aim to survey only small regions around the Galactic centre and other small regions with scientifically interesting sources. 
Small-JASMINE will have a very bright star limit of Hw=11.5 mag (Hw-band:1.1 $\sim$ 1.7 $\mu$m) with accuracies comparable to Gaia but 
lack the all-sky coverage and faint star limit that only GaiaNIR can achieve. The Euclid mission will have high quality NIR photometry
but its science cases are focused on the extragalactic sky which is presently defined by the regions covering $\left|b\right|>30^\circ$ which precisely
avoids those regions GaiaNIR would be most interesting for.

\subsubsection{The spiral arms}\label{spiralarms}

It is surprising that very little is known about the spiral structure of the Milky Way. The Bar and Spiral Structure Legacy (BeSSeL) 
survey\footnote{See this \href{http://bessel.vlbi-astrometry.org}{link}.} and the Japanese VLBI Exploration of Radio Astrometry (VERA) 
survey\footnote{See this \href{http://veraserver.mtk.nao.ac.jp}{link}.} have yielded over 100 parallax measurements in the spiral arms and the 
central bar with typical parallax accuracy of about $\pm$20 $\mu$as, and some as good as $\pm$5 $\mu$as. 
These radio measurements are providing good constraints on the fundamental parameters of the Galaxy, including the distance to the Galactic 
centre. Gaia on the other hand is very limited by extinction at optical wavelengths and will not be able to freely probe the Galactic plane \citep{2014ApJ...783..130R}.

Spiral arms are the main areas of star formation in the Milky Way, and are responsible for a significant portion of radial migration and disk 
heating. A greater understanding of their structure will help in understanding these processes.
By incorporating an NIR capability a new Gaia-like mission would be able to see through dust and would allow us to better map the spiral 
arms and the star forming regions, not just for a few hundred objects as is the case for radio astrometry, but potentially for hundreds 
of millions of objects, revealing much more detailed structure that radio surveys cannot hope to achieve. A new Gaia-like mission 
going into the NIR could provide model-independent distances and proper motions, avoiding the need to use extinction maps together 
with galaxy modelling. Galactic archaeology relies on accurate distances and needs better accuracy towards the spiral arms and the 
galactic centre to give accurate stellar ages and astrophysical parameters in these important regions.

\subsubsection{Galactic rotation curve and Dark Matter}\label{rotation}

Currently, the gravitational force at different points in the inner disk of the Milky Way is not well known. 
This property is key to learning about the distribution of dark matter near the centre of the Milky Way 
(and indeed luminous matter). GaiaNIR will be the only instrument able to solve this problem since it will 
unveil the inner dynamics of the disk, in unprecedented detail, from hundreds of millions of stars, in the 
bulge, bar, spiral arms, and between the spiral arms. This will allow us to perform high spatial resolution 
mapping of the dark matter distribution in these regions. This will be key to resolving questions regarding 
the nature of dark matter particles, by showing us whether the Galaxy has a cored or cusped dark matter halo, 
whether there are any thin, disc-like components to the dark matter distribution, and whether spiral arms 
have their own dark matter component.

In external galaxies, and the outer parts of our own (where we can assume axisymmetry), it is common to approximate 
the gravitational force in the plane using a rotation curve. In the inner Milky Way, this is not an appropriate 
approach because of the non-axisymmetry and because some key tracers (H$_{\rm I}$ or CO regions) provide poor 
spatial coverage \citep{2015A&A...578A..14C}. A more sophisticated approach is required\citep[e.g.][]{2015MNRAS.448..713P} 
along with more data. VLBI cm measurements of masers \citep{2014ApJ...783..130R} are excellent probes of the rotation curve, 
kinematics and structure in the low latitude regions of the Milky Way. Unfortunately they are too limited in the number 
of targets (at most a few hundreds when their survey is complete), and are only sensitive in star forming regions (thus only 
in the spiral arms or along the dust lanes of the bar), to provide high spatial resolution. VLBI mm measurements with 
ALMA will also play a role in a very near future to do similar science as cm-VLBI, likely with many more detections 
because there are more emission lines at mm- than at cm-wavelengths. Gaia and VLBI will help in the quest to understand 
the inner dynamics of the Galaxy but an NIR capable astrometric mission would allow much greater insight and may resolve 
these issue.

The incredible strength of GaiaNIR with respect to radio interferometry is that it will not be restricted to any particular disk
perturbations or regions and will observe millions of sources. GaiaNIR will be the only way to probe both the low latitude arm
and inter-arm regions, as well as directly inside the bar. It will improve greatly on cm- and mm-VLBI measurements restricted to
star forming regions only, and 21~cm H$_{\rm I}$ or mm CO measurements which do not allow direct measurements of distances and 
proper motions of gas clouds. GaiaNIR is the only way to determine hundreds of millions of velocities at all points in the
Galactic plane and model their dynamics, even at $z\rightarrow 0$kpc. This will make it possible to make a comparison with 
stellar rotation curves of other galaxies, and the cusp-core controversy (among others) will then be accurately investigated.

\subsubsection{Clusters}\label{clusters}

The formation and evolution of clusters, which are the source of the stellar field populations, are key to improving our knowledge on the birth of stars.
Clusters are often located in the spiral arms of the Milky Way; they are composed of young stars that have recently formed 
in the disk. The stars belonging to a cluster have roughly the same age and metallicity, having formed from the same cloud of gas and dust,
and can be used to probe the galactic disk structure and formation rate, as well as to study the young star properties and their formation process
 \cite[see for example:][]{2015ApJ...812..131K}.

The internal dynamics of embedded clusters are needed to study the small scale structure of the molecular clouds
and shed light on the conditions and physics of the cluster at the epoch of star formation. Likewise the bulk dynamics of these clusters
allow one to probe the large scale structure of the molecular clouds and the kinematics tell us about the state of the gas at the epoch of formation.
An NIR option would allow the dusty star forming regions to be globally surveyed for the first time. These important stellar birth places 
are in the regions where extinction intervenes to make optical observations, such as those of Gaia, difficult.

On the other hand open clusters are excellent tracers of the Galactic potential if their tangential motions and accelerations are accurately measured,
particularly as the cluster distances can be very accurately constrained from astrometry and photometry.  Accurate proper motions from combined missions
would allow us to derive cluster membership more clearly. This is already partially possible with the Gaia data alone but a combined mission separated by 
20 years would allow internal accelerations and the initial mass functions to be studied and allow us to probe much larger volumes of the Galaxy. 

ALMA, ALMA-VLBI, and the SKA will help us to understand the way gas clouds evolve (condense, fragment) and form stars with unprecedented 
detail. However, it will not be possible to perform astrometry on a large-scale with these instruments.
Gaia-NIR is an excellent instrument to complete ALMA, as it will enable us to reveal the position-velocity 
phase space of, for example, low mass, newly born stars, etc. GaiaNIR will be a unique opportunity to link/compare the velocities of these new 
stars to the dynamics of the gas, thus fully understanding the star formation processes. GaiaNIR will also help to constrain the distances 
of any gas clouds whose line-of-sight kinematics are similar to nearby stars.

Finally, 3-D maps of the ISM need good distances and, in principle, a solution can be found for the extinction. 
Gaia can only do this up to $\sim$~1~kpc but going further into the NIR would greatly help.

\subsection{The halo, streams, hyper-velocity stars and the local group}\label{halostreams}

Gaia is a fabulous mission for detecting and characterising nearby streams that cross the disk of the Milky Way but it will not be sufficient 
to discover and characterise most of the stream-like structures in the halo. However, improvements in the accuracy of proper motions would 
allow a new mission to resolve tangential motions in streams and local dwarf galaxies, with a potential accuracy of 2--3~km~s$^{-1}$ 
for specific samples out to $\sim$100~kpc. This will provide great insight into the  gravitational potential in the outer reaches of 
the Milky Way. This is only possible by exploiting the long time baseline allowed by combining Gaia measurements 
with those from a future astrometric mission.

The total mass of the Milky Way is an important quantity to determine because it tells us where the Milky Way 
fits in a cosmological context. The local group has been called the ``Ultimate Deep Field" \citep{Boylan-Kolchin19062016} 
because it probes a large co-moving volume at high redshift, and can be observed with great detail. Important 
questions in cosmology such as the ``Missing satellite" problem and the ``Too Big to Fail" problem are related to whether the 
Milky Way is typical of galaxies seen in cosmological simulations. To make this comparison in a fair way one needs to 
know the mass of the Milky Way.

\subsubsection{The halo and streams}\label{halo}

Streams in the Milky Way halo are formed when satellite galaxies or globular clusters are pulled apart in the tidal 
potential of the Milky Way. The stars then drift apart because they are on different orbits and form a (typically thin) 
band of stars across the sky. This makes them sensitive probes of the Milky Way's potential because the model of their 
formation is relatively straightforward, and their narrow distribution across the sky provides tight constraints in 
some ways (though ones that require more phase-space information if they are to be useful).  This will allow us to determine 
the dark matter distribution at large radii, including any flattening of the potential, and the total mass of the Galaxy. 
A future astrometry mission would provide highly accurate proper motions and more accurate distances to these stars, 
which will allow much more precise determination of the potential of the Milky Way at these large radii.

A further exciting opportunity opened up by improving proper motion measurements for stars in Milky Way streams is that by 
finding ``gaps" in the streams (more easily found if we have proper motions as well as positions for stars), we might see the 
influence of dark matter sub-haloes in the Milky Way's halo \emph{even though they contain no stellar matter}. It is hypothesised 
that such sub-haloes are common in the Milky Way halo, but the only way to detect them is through their gravitational interaction 
with luminous matter. When one of these dark matter haloes has a flyby encounter with a stellar stream, it can create a disturbance 
(or gap) that may be detectable with sufficiently accurate proper motion measurements \citep{2016PhRvL.116l1301B,2012ApJ...748...20C,2016MNRAS.457.3817S}.
The dynamically cold nature of the stellar streams makes them ideal candidates for detecting these flybys.

Another method of determining the mass of the Milky Way out to large radii is by modelling the motion of a tracer population in the
outer halo that can be assumed to be in equilibrium. The major difficulty with this endeavour has historically been that there is a 
degeneracy between the mass enclosed within a given radius and the orbital anisotropy of the tracer population \citep{2006MNRAS.369.1688D}. 
This is only a problem because the only available measured velocity component was the line-of-sight velocity, and for stars in the outer 
halo this component is nearly identical to the Galactocentric radial velocity. The availability of accurate proper motions for these 
stars will break this degeneracy and enable us to learn the mass of the Milky Way from the tracers we see.

Further, it is an open question what fraction of the stellar halo is in substructures and what fraction is in a smooth component. Sometimes 
substructure can be found as an over-density on the sky, but the availability of accurate proper motions will make it easier to find 
substructure in the larger phase-space.

All of the above questions cannot be answered by Gaia alone due to the low accuracy of the proper motions at great distances ($\sim$10~kpc).
However, a new mission combined with the older Gaia/Hipparcos/Tycho-2 catalogues would give the much longer baseline needed to get very 
accurate proper motions; remember a factor of 14 better in the two proper motion components (see section \ref{Measurement}) 
to open up these critical areas to understand the Galaxy in a much grater volume than is currently possible.

\subsubsection{Hyper-velocity stars}\label{hypervel}
\cite{2015ARA&A..53...15B} and references therein, have shown that the origin of hyper-velocity stars (HVSs) are most 
likely due to gravitational interactions with massive black holes due to their extreme velocities. Future and very accurate
proper motion measurements are a key tool to study these objects. Precise proper motion measurements, due to the largely radial 
trajectories of these stars, combined with radial velocities can provide the three dimensional space velocity of these objects. 
Unfortunately, known HVSs are distant and on largely radial trajectories \citep{2015ARA&A..53...15B}. Some HVSs originate in the 
Galactic centre while others have an origin in the disk but an origin in the Magellanic Clouds or beyond is also possible. 
Proper motion accuracies from Gaia (or combined missions) are needed to reconstruct their trajectories and distinguish between 
the different possible origins.

\cite{2005ApJ...634..344G} have shown that precise proper motion measurements would give significant constraints on the structure (axis ratios and 
orientation of triaxial models) of the Galactic halo. Triaxiality of dark matter halos is predicted by cold dark matter models of galaxy formation 
and may be used to probe the nature of dark matter. Using data from Gaia--GaiaNIR and combined with distances and radial velocity measurements
would allow for a factor of 14 improvement in the accuracy of the proper motions to further constrain our understanding of HVSs and the 
Galactic potential. Adding the NIR capability would also allow us to probe more deeply into the Galactic centre and to potentially detect small 
populations of HVSs closer to their ejection location.

\subsubsection{Local group}\label{localgroup}

Astrometrically resolving internal dynamics of nearby galaxies, such as M31, dwarf spheroid galaxies, globular clusters, the Large and Small Magellanic Clouds (LMC, SMC), sets requirements on the accuracy. For example, the LMC has a parallax of 20~$\mu$as and an accuracy of about 10\% is needed, which 
is just within the reach of Gaia. Precise mapping of dark matter (sub-)structure throughout local group and beyond is possible with accurate proper motions.
Gaia will not be able to directly measure internal motions of the nearby galaxies. However, by combining Gaia proper motions with a new Gaia-like mission
opens up the new tantalising possibility of measuring their internal motions and thus astrometrically resolving the dynamics within the Local Group.
A number of science cases could be within reach:
\begin{itemize}
\item Using photometric distances to the LMC allows us to learn about internal dynamics and structure. 
\item Mapping the ISM in the LMC would also give new insight.
\item Using proper motions to determine rotational parallaxes to M31, M33 and other local galaxies.
\item Dynamical measurements of the mass distribution of the Milky Way and M31
\item Probe the internal kinematics classical dwarf spheroid galaxies. 
\item Mapping the dark matter sub-structure throughout the local group out to M31
\item Resolve the core/cusp debate\footnote{The discrepancy between the observed dark matter density profiles of low-mass galaxies and the density profiles predicted by cosmological N-body simulations.} with 6D phase space information.
\item Ultra-faint satellite galaxies would be accessible and their orbits could be determined.
\end{itemize}

Dwarf spheroid galaxies (dSph) are fascinating systems and likely fossils of the re-ionisation epoch\footnote{This section is extracted from an unpublished note from Rodrigo Ibata, see this \href{https://dl.dropboxusercontent.com/u/49240691/GaiaIbata.pdf}{link}.}. 
In order to study their internal proper motions there is a need to disentangle the Milky Way environment using very accurate 
proper motion measurements.  The dSphs are small, almost spherical, agglomerates of stars that orbit more massive hosts. 
In the local group these dSph galaxies are the most numerous class of galaxies, with 15 discovered around the Milky Way, 25 
around the Andromeda galaxy, and possibly one orbiting M33. Several other nearby galaxies have also been found to harbour 
these satellites. It is believed that hundreds of dSphs and globular clusters were accreted during the formation 
of the Galaxy, and their remnants should be visible as coherent phase-space structures. The challenge 
is of course to identify the stars with common dynamical properties.

An interesting aspect of these galaxies is their identification as possible cosmological “building blocks” of the type predicted 
by Cold Dark Matter (CDM) theory. A generic feature of a Universe in which the dominant form of matter is in the form of 
collisionless (i.e. cold) Dark Matter (DM) is that galaxies should be surrounded by tens of thousands of dark matter satellites 
with masses comparable, or larger, than those of dwarf galaxies. While the number of observed satellites is very much lower than 
this, it has often been suggested (or assumed) that baryons collapsed to form substantial stellar populations in only the most 
massive of the dark satellites. Other dark satellites may have lost their baryons, or had star-formation suppressed for a variety 
of reasons. Thus the dwarf satellite galaxies may represent the surviving remnants of the original population of primordial galaxies 
that coalesced and merged to form our Milky Way.

To confirm this picture we need to understand the internal dynamics of the dwarf galaxies, and to date, the necessary observations 
to test these ideas have remained beyond our reach. The issue is that there is a trade-off between dark matter content and tangential 
anisotropy and/or tidal heating. A system with stars on preferentially tangential orbits to the galaxy and that has also been heated by 
repeated tidal encounters with the Milky Way can masquerade as a highly dark matter dominated galaxy. The only way to be sure of 
the true dark matter content is to measure the internal proper motions of the stars in the dwarf galaxy, not only the bulk proper 
motion of the system (which will be feasible with Gaia). Only a joint mission in the future can hope to achieve this challenging goal.

\subsection{Low mass brown dwarfs and free floating planets}\label{browndwarfs}

A strong limitation of Gaia is that it observes in the optical. As such, it is blind to faint very red objects and sources 
in extinction regions.  Examples of such objects are: brown dwarfs, which are substellar objects that occupy the mass 
range between the heaviest gas giants and the lightest stars (roughly between 13 and 80 Jupiter masses); red dwarfs which are small and 
relatively cool stars on the main sequence, either K or M spectral type with masses from 0.075 to about 0.50 solar masses. 
Another example is free-floating planets which are interesting to study to determine if they were ejected from planetary systems or 
formed in collapsing dust clouds in a similar manner to stars. Looking for these objects with Gaia in open clusters and star forming 
regions will  remain a challenge and will only be possible in nearby regions with low extinction.
On the other hand an NIR capable mission would allow such objects to be detected within a large volume but also crucially in the extinction 
(open clusters and star forming) regions that are of great interest and would shed light on the Initial Mass Function (IMF) of these 
regions which is a topic in astronomy that initiates vigorous discussion \citep[see for example][]{2010ARA&A..48..339B}.

A key issue in state-of-the-art brown dwarfs and free floating planets research is in the theoretical evolutionary models, which are 
needed to assign mass given a measured luminosity and an assumed age. The biggest source of uncertainty in these models is in the 
initial conditions: at what level the reservoir of heat starts that sets the initial brightness of the cooling curve. In order to 
discriminate between models, observations in clusters and associations with well defined ages are needed, i.e. Hyades, Pleiades, 
Alpha-Per, Praesepe, etc. all of which Gaia can reach for normal stars, but not for brown dwarfs and free floating planets. 
While it is true that brown dwarfs and free floating planet members can be identified in extant IR proper motion surveys and we 
can use the Gaia measured mean cluster distance, that ignores depth effects, i.e. blurs the resulting HR diagram and prevents 
critical comparison with the models and discrimination between them. GaiaNIR would solve the problem by measuring the distances 
directly and accurately for many nearby clusters and associations yielding sharp observational isochrones with which to confront 
evolutionary models with different initial conditions.

\subsection{White dwarfs}

White dwarfs (WD) are the final evolutionary stage of intermediate and low mass stars (about 95\% of all stars end as WDs remnants).
Their study provides key information about the late stages of the star's life, and also of the structure and evolution of the Galaxy
because they have an imprinted memory of its history \citep{2001ASPC..245..328I, 2005ApJS..156...47L}. 
Through comparison of the empirical and theoretical Luminosity Functions (LF) of WDs one can derive the age of the Galaxy and its
star formation rate. The LF allows the reconstruction of the IMF. Specially important are the oldest members of the WD populations 
of thin and thick disks and halo, which were formed by high-mass progenitors and evolved very quickly to the WD stage, being very
cool and faint in our present days.
An NIR facility will allow to characterize the cool WD population much better than Gaia in the optical, which has limited
capabilities \citep{2014A&A...565A..11C} both in terms of detection and parametrization. As an example, pure-H and pure-He
atmospheres can only be distinguished in the NIR regime.

\subsection{Astrometric microlensing}

Black holes, neutron stars, brown dwarfs are expected to fill our Galaxy in large numbers, however, such dark objects are 
very hard to find, as most of them do not emit any detectable light \citep{2016MNRAS.458.3012W}. About thirty years ago 
Bohdan Paczy{\'n}ski \citep{1986ApJ...304....1P} proposed a new method of finding those compact dark objects via photometric 
gravitational microlensing. This technique relies on continuous monitoring of millions of stars in order to spot 
its temporal brightening due to space-time curvature caused by a presence and motion of a dark massive object. 
Microlensing exhibits itself also in astrometry, since the centre of light of both unresolved images 
(separated by $\sim$1 mas) changes its position while the relative brightness of the images changes in the 
course of the event. Astrometric time-series at sub-mas precision over a period of several years would provide 
measurement of the size of the Einstein Ring, which combined with photometric light curve, would directly 
yield the lens's distance and mass. Most microlensing events are detected by large-scale surveys, e.g., OGLE and, 
in future possibly also the LSST.  

At typical brightness of V=19-20 mag Gaia-like missions would be capable at providing good-enough astrometric 
all-sky follow-up of photometrically detected microlensing events. Gaia will provide superb astrometric time 
series and once combined with parallax measurements, a mass and distance of the 
lens will be measured without the need of any assumptions about proper motions. A NIR capability would be best 
for Bulge microlensing - lower extinction, more sources to monitor, etc. \citep[see for example][]{2009MNRAS.396.1202K}.
Detection of isolated black holes and a complete census of masses of stellar remnants will for the first time 
allow for a robust verification of theoretical predictions of stellar evolution. Additionally, it would yield 
a mass distribution of lensing stars as well as hosts of planets detected via microlensing. 

\begin{figure}
	\begin{center}
		\includegraphics[width=13cm]{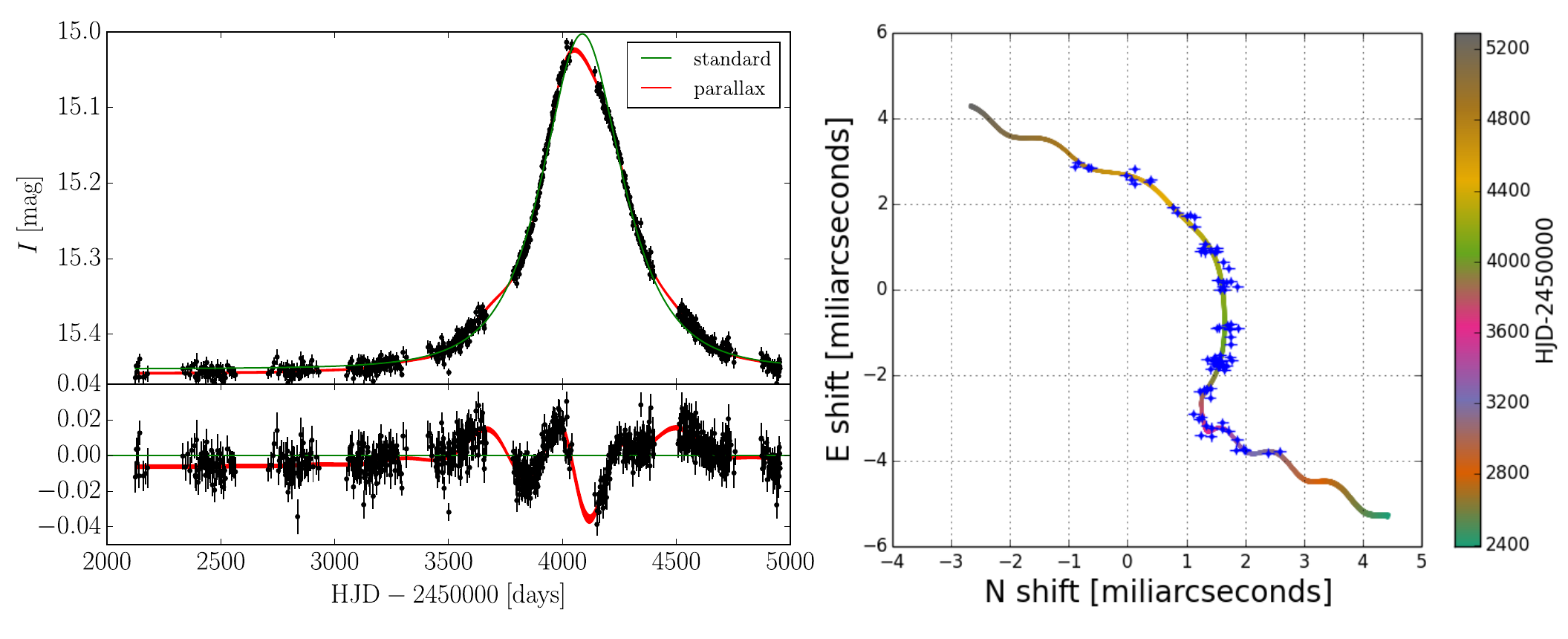}
		\caption{Microlensing event, OGLE3-ULENS-PAR-02, the best candidate for a $\sim$10M$_\odot$ single black hole. Left: photometric data from OGLE-III survey from 2001-2008. Right: simulated astrometric microlensing path for a similar event.}
	\end{center}
\end{figure}

\subsection{Prospects for exoplanets and long period binaries}\label{exoplanets}

Precision astrometry in principle allows to detect planet-induced stellar motions
and to detect long period (up to 40 years) binary systems. The orbital
elements of these systems can be derived, including the inclination,
even for multiple-planet systems (see, for a real application,
\cite{2010ApJ...715.1203M}).  The required astrometric precision is,
however, challenging, and it is only with Gaia that astrometric
detection and orbital element fitting are about to become routine.

An NIR mission would allow exoplanet detection around cooler, lower
mass stars than those observed by Gaia. The astrometric signal for a
body of mass $M_b$ orbiting a star of mass $M_*$ goes as $M_b M_*^{-1}$,
therefore astrometric detection becomes easier for planets around
low-mass stars, or for stellar binary systems where the two components
have comparable mass. This means that planets would become detectable
with lower masses than those accessible by Gaia (e.g., Neptune-like
planets). Binary stars, which have larger astrometric signatures than
planets, are expected to be detectable to much larger distances
(in theory up to $\sim$10~kpc), being mostly limited by their magnitude.

The new astrometric mission would allow the detection of planets with
significantly longer periods than by Gaia alone. Our tests (Ranalli et
al., in preparation) and previous literature
\citep{2014ApJ...797...14P,2008A&A...482..699C} all agree that the
longest detectable period using the astrometric method roughly
coincides with the mission lifetime. This sets a maximum period
$P\sim$5~years for Gaia in its nominal mission, or $P\sim$10~years if an
extension is granted. A new mission launching approximately in 2035
and using a similar scanning law to Gaia's would put the total time
span in the $\sim$26--31~year interval, depending on the new mission
lifetime. For comparison, the planet Saturn has a period of 29~years. 
This means that the combination of the two missions will allow the
detection of gaseous giants orbiting their stars at approximately the
same distances as they are in the Solar System. A similar limit will
hold for long-period binary stars: if the period is 30~years and both
have masses 0.5~$M_\odot$, the semi-major axis will be 12~AU (slightly
larger than the Saturn orbit). Since binaries have much larger
astrometric signatures (and hence, better signal/noise ratios), it may
be possible to detect them even if their period is somewhat longer
than the limit above, possibly to $\sim$40~years.
\begin{figure}[tbh]
	\begin{center}
		\includegraphics[scale=0.4]{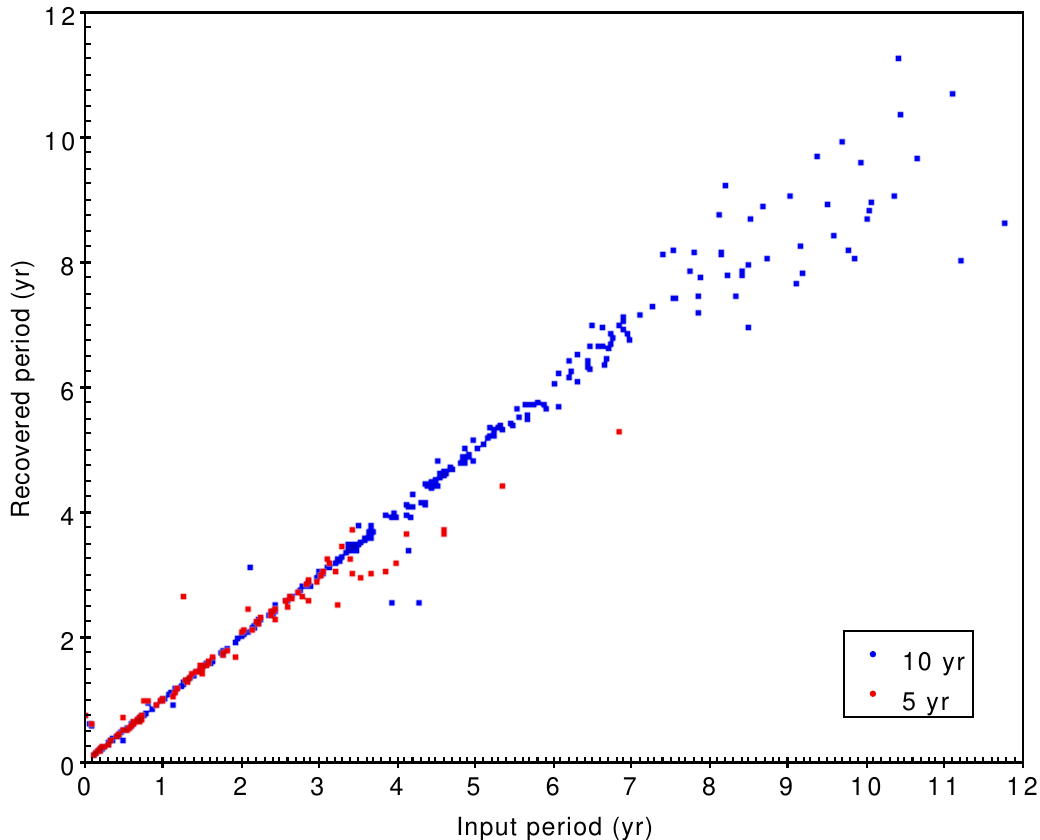}
		\includegraphics[scale=0.4]{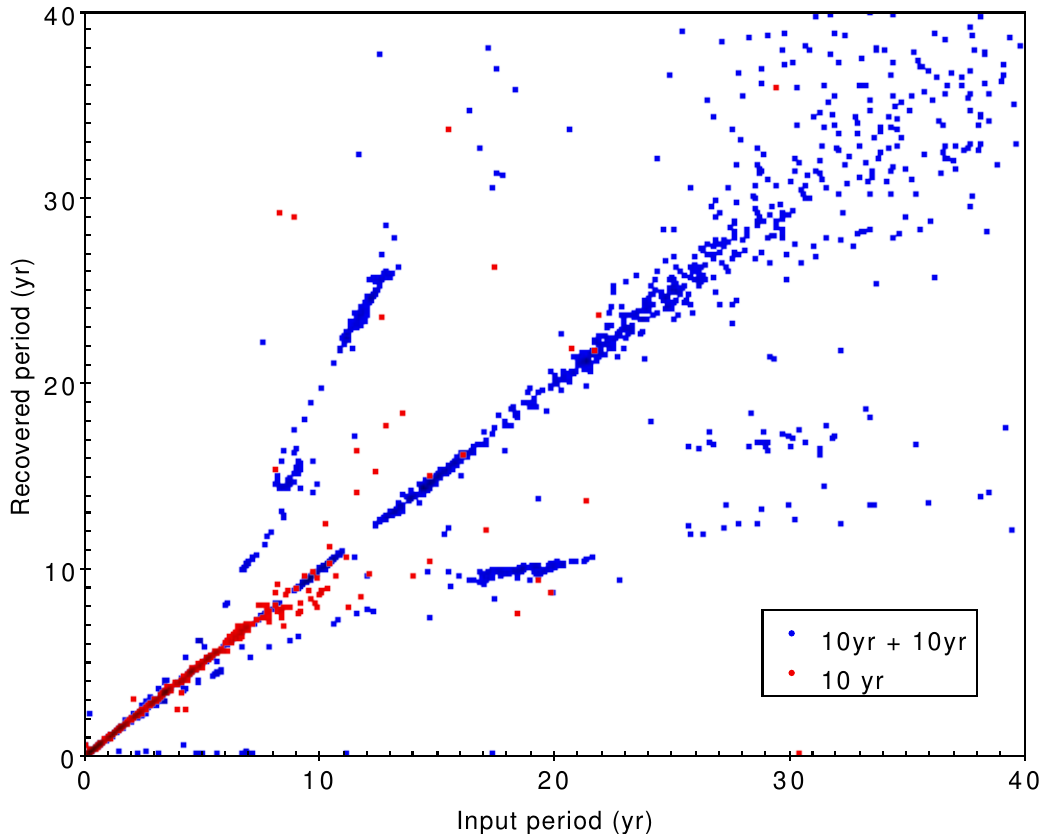}
		\caption{\em{The left figure shows the improvement in the possibility of detecting planets and recovering their periods when going from the nominal Gaia mission (5~years) to an extended mission (10~years). While the nominal mission allows to detect planets up to $\sim$4~years, the extended mission shifts the limit around $\sim$10--11~years.  The right figure shows the improvement when adding a second 10~year mission with launch in 2035 to data from an extended Gaia mission (10~years).
		Planets with periods up to 30~years may be reliably detected. In some cases also planets with periods up to 40~years may be detected, though with larger deviations on the periods (Images Ranalli et al., Lund Observatory, in preparation).
		}} \label{Fig:exoplanets}
	\end{center}
\end{figure}

Proper motions derived from short missions can be affected by systematic errors 
due to the motions in unresolved binaries and the error depends on the mass and orbital 
period of the pair. Such systems are poorly surveyed at present and could lead to the 
discovery of Solar System analogues. The unresolved double stars can be detected due 
to the large residuals in the astrometric solution, which is normally based on 
linear motion for a single star. Additionally, a comparison of the proper motions from 
each mission with a joint solution from both missions can improve binary population 
statistics and reveal the acceleration in the orbit. More detailed non-linear modelling 
for multi-star systems is then needed for the classification of such binaries and this 
is greatly enhanced by the second mission and the longer time baseline leading to new 
discoveries.

Tests also show that the presence of a long gap ($\sim$11--16~years)
in the data collection does not introduce too large biases in the
recovered orbital elements. Aliases have been found to be easily
identifiable; and their presence is greatly mitigated when at least
one mission is granted an extension. With two short missions (5+5~year)
we expect fractions of false positive and false negative detections
under $\sim$10$\%$; with two long missions (10+10~year) the fractions
would be significantly lower. Also, planets and far stars would become
detectable with lower signal/noise ratios than in the case of a single
mission. 

A space-based proper motion catalogue derived from two 
Gaia-like missions, 20 years apart, for 1 billion stars would supersede any 
future ground based catalogue. Synergies between the two Gaia-like missions 
and the Plato mission would allow a comprehensive comparison of the short period 
planets detected by Plato and the long period planets from the Gaia-like missions.
This would be done for common systems giving quantitative insight into possible 
Solar System analogues.

\subsection{Maintenance of the celestial reference frame}\label{refframe}

The Gaia optical reference frame will degrade over time. As an example, if the reference frame is accurate to 0.4 mas 
(at G = 20) at its reference epoch it will degrade to 14 mas after 50 years.
Dense and accurate reference grids are needed for forthcoming Extreme, Giant and Overwhelming telescopes but also 
for smaller instruments currently operating or being planned. An important aspect of reference frames is to link 
them, cross-matching with absolute coordinates, to reference frames at other wavelengths to produce reference 
grids for various surveys. This requires the maintenance of the accuracy of the Gaia optical reference frame at 
an appropriate density that is useful for new surveys. A key improvement is to expand the Gaia optical reference to 
the NIR increase its density in obscured regions and to then link this new reference frame to the ICRF.

The reference frames are based on quasars, which are not point sources and have astrometric variability at the 10~$\mu$as 
level due to internal structure and activity. Maintaining the optical/NIR reference frame is a vital service to the community 
and is not feasible from the ground. Celestial reference frames are important for a number of reasons:
\begin{itemize}
\item To refer positions of fixed or moving sources
\item To detect tiny motions (e.g. exoplanet detection by relative astrometry)
\item To quantify the motion of sources without any kind of bias:
\begin{itemize}
\item Motion of stars in the Galaxy
\item Differential rotation of the Galaxy
\item Dynamics of star clusters
\item Investigate rotational and translational motion of external galaxies
\item Dynamics of the Solar System
\end{itemize}
\item Source cross-identification in $\gamma$, X-ray, visible, IR and radio wavelengths
\item To monitor the rotation of the Earth and study tectonic plate motions
\item Angular positions (and distances) of quasars, galaxies, stars, planets, spacecraft navigation, GNSS maintenance, etc.
\end{itemize}

The need for a global survey mission like Gaia to maintain the realisation of the celestial reference frame is a 
science objective in itself as it lies at the heart of fundamental astrometry and provides a reference grid for 
much of modern observational astronomy.
The Gaia celestial reference frame will provide a realisation of the reference frame with  bright quasars as a dense 
optical version to G $\sim$20.7. This frame will degrade with time but will immediately supercede any other 
optical reference frames and will be the standard optical reference frame for the astronomical community for many years.
The Gaia frame will be compatible with ICRF-2 (and ICRF-3) --  within the uncertainty of the latter. There are six degrees 
of freedom  the orientation $\vec{\epsilon} = (\epsilon_x,\ \epsilon_y,\ \epsilon_z)$ and the spin 
$\vec{\omega} = (\omega_x,\ \omega_y,\ \omega_z)$. The orientation is determined using a set of a few thousand radio 
VLBI objects with known positions and proper motions in the ICRS independent of Gaia while the spin is determined using 
a set consisting of hundreds of thousands of quasar-like objects ($10^5$--$10^6$) taken from ground based and photometric surveys.
Eventually, the reference frame will be uniquely defined by Gaia itself using quasar classification and this will also hold for 
future missions.
\begin{center} 
	\begin{figure}[tbh]
		\centerline{\includegraphics[scale=0.29]{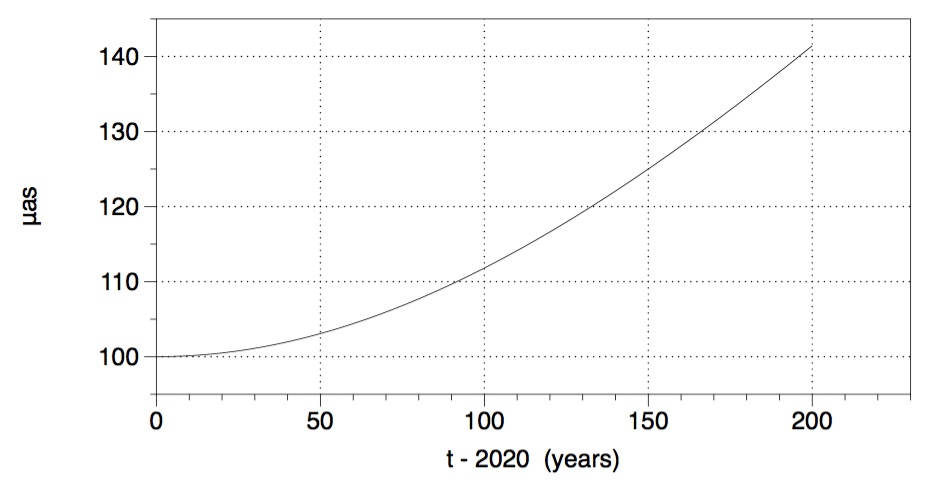}
			        \includegraphics[scale=0.23]{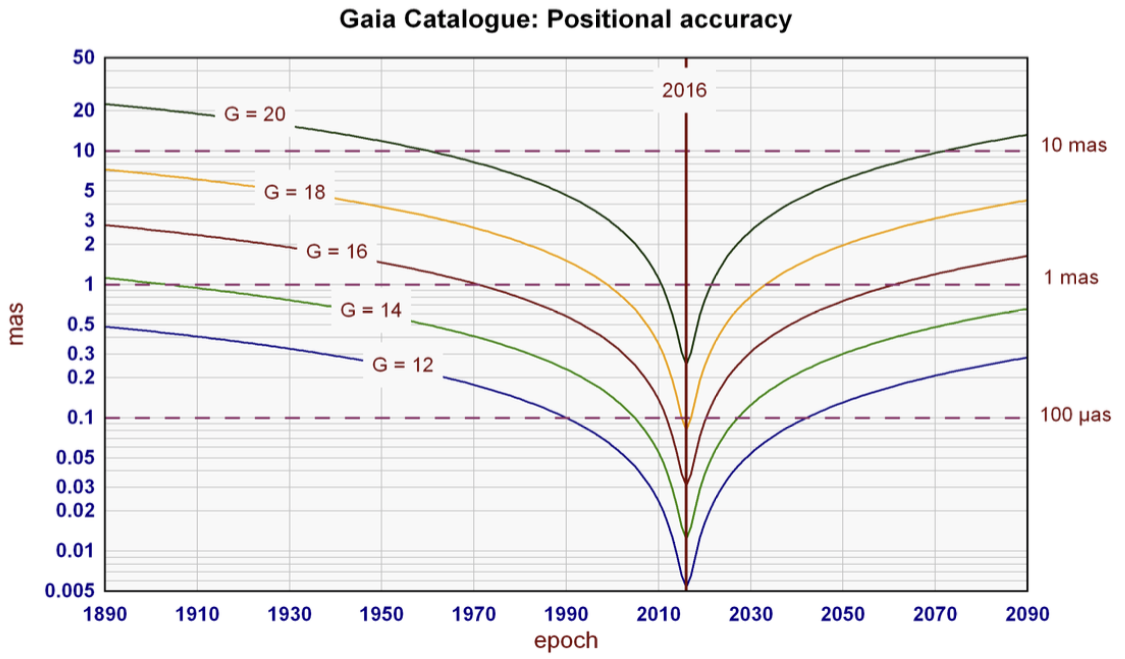}}
		\caption{\em{Left image shows the degradation of the astrometry of the individual sources in the Gaia-CRF while the right image shows the positional accuracy of the Gaia catalogue over time. (Image F. Mignard).
		}} \label{Fig:Degradation}
	\end{figure}
\end{center}
	
The accuracy depends on radio-optical offsets which are largely unknown at the level of Gaia--ICRF accuracies.
These radio-optical offsets are not well founded on observational evidence but could range from 0.1 to 1 mas for some sources.
Their impact on the Gaia reference frame has recently been assessed \citep{2014ApJ...789..166P, 2016A&A...589A..71B} but should 
be randomly oriented and act only as an additional noise. The Gaia frame has a reference epoch but the accuracy of the frame will 
degrade with time primarily because of the uncertainties in the proper motion components but also because of systematic proper motions 
resulting from the acceleration of the Local Group. The quasars also have a systematic proper motion pattern of $\sim$4.3~$\mu$as~yr$^{-1}$, 
due to the acceleration of the Solar System towards the galactic centre (see also Sec. \ref{patterns}). 
This acceleration can be measured and the reference system maintained with an accuracy $< 0.5~\mu$as~yr$^{-1}$. 
Individual positions of the primary sources will have an accuracy of $\sim$100~$\mu$as and will degrade very slowly 
(see Figure~\ref{Fig:Degradation} left) but other transverse motions will show up over time. For example, cosmological proper motion 
is the instantaneous velocity of the Solar System with respect to the cosmological microwave background (CMB) (see also Sec. \ref{patterns}). 
This will cause extragalactic sources to undergo an apparent systematic proper motion. The effect is referred to as cosmological or 
parallactic proper motion (Kardashev 1986). 

In conclusion, a new Gaia-like mission will be necessary in the coming decades in order to maintain the optical realization of the 
celestial reference frame. This on its own is a strong science case due to the need to maintain dense and accurate reference grids
for observational astronomy in the optical. However, the addition of NIR astrometry will increase the density of this grid in 
obscured regions and provide a link to the ICRF in a new wavelength region.

\subsection{Stars}\label{stars}

High-precision astrometry in the NIR will have an impact on a wide variety of stellar physics topics. 
Here, we highlight three areas related to stars that will profit from advances in NIR astrometry technology: 
star clusters and associations, binary and multiple systems, and stellar standard candles.
A common aspect applying to all of these areas is that at NIR wavelengths we will be able to reach deep into 
those regions of the Galaxy which are obscured by dust clouds. These regions in the Galactic plane and near 
the centre of the Galaxy are particularly interesting for studies of the conditions of star formation, as 
they are the common birthplaces of stellar clusters and associations.

%
%
Stars are continuously formed in groups of tens to thousands and evolve together for a shorter ($\sim$100~Myr) 
or longer time (a few Gyr) in associations or open clusters, respectively, depending on whether they are 
gravitationally unbound or bound. These clusters are a unique laboratory for observing the outcomes of star 
formation and evolution. More than 2000 open clusters are known today, most of them within a distance of 
$\sim$2~kpc from the Sun \citep{2013A&A...558A..53K}. 
This is on the order of 1\% of the total population of over $10^5$ open clusters expected in the Galaxy 
\citep{2016arXiv160700027M}. 
The Gaia mission will extend the census of open clusters to $\sim$5~kpc, potentially increasing the 
number of mapped clusters to $\sim$10\% of the total. However, most of these will be located at high galactic 
latitudes with low interstellar extinction.

A mission with 14 times better accuracy and NIR capabilities would cover half of the Galaxy or more, including 
regions towards the Galactic centre. It would thus enable us to probe a much more diverse range of environments 
for cluster formation, in terms of stellar and gas density, and metallicity (all of which increase towards 
central regions). Cluster properties such as stellar density (cluster size and number of members), dynamics, 
and age as a function of location within the Galaxy will provide strong constraints to models of star formation. 
Furthermore, the range of stellar types will be increased to include low-mass stars (subsolar) whose fluxes peak 
in the NIR. This will allow a better characterisation of the IMF at low masses and its 
dependence on the environment, and will again provide clues on the physics of star-forming processes. Finally, 
with an increased astrometric accuracy it will be possible to access the more crowded inner regions of each cluster.

%
%
Another area which will profit from optical/NIR astrometry is the study of binary stars and multiple systems. 
The measurements of astrometric orbits and accurate distances will make it possible to determine accurate masses 
of the components of binary systems, which are crucial to advancement in several areas of stellar physics.
For a planetary system, the mass of the host star must be known in order to determine the masses of the planets. 
Masses of single stars are determined with the help of stellar evolution models, and have uncertainties of up to 30\%, 
leading to 15\% uncertainty in planet masses (see e.g. the exoplanets.org database). This is particularly true for
low-mass stars, which also seem to be the most-frequent planet hosts \citep{2015ApJ...807...45D}. 
Masses of binary stars determined from astrometric orbits together with observables to be compared with 
predictions from stellar models provide the necessary constraints for developing the most realistic models. 

More exotic objects which will benefit from improved mass estimates from binary orbits are neutron stars 
and stellar-mass black holes. This science case is similar to one of those envisaged for the SIM astrometry 
mission \citep[see][and references therein]{2008PASP..120...38U}. 
The equation of state for matter at densities beyond those of nuclei is not known. Several proposals exist, 
with different predictions for neutron star masses, and current mass estimates are compatible with all of those.
Likewise, large uncertainties are associated with current mass estimates for Galactic black holes and hamper 
our understanding of their nature.
In addition, dynamics of neutron star or black hole binary systems measured from accurate proper motions will 
allow to constrain their formation mechanism.

%
%
Optical/NIR astrometry will enable extensive local tests of stellar standard candles \citep[e.g.][]{2008PASP..120...38U}. 
Accurate distances of Cepheids and RR Lyrae stars throughout the Galactic plane, including the Galactic 
bulge, will result in ultimate period-luminosity relations, since the key uncertainties of variable extinction and 
metallicity are significantly reduced in the NIR \citep[see Figure 15 in][]{2016A&A...591A...8A}. Through an extension of the measurements to local group galaxies, 
the validity of these relations can be tested for different chemical environments and galaxy types. 
For Cepheids, a better understanding of the physics of the pulsation mechanism may improve the applicability 
of the period-luminosity relations for determining accurate galaxy distances. Advancement in pulsation models 
requires accurate masses of Cepheids, which can be obtained from astrometric orbits of binary systems with Cepheid components.
\cite{2011A&A...530A..76W} have shown that the Cepheids are 
very difficult to observe towards the galactic centre (see Fig. 4) and at low latitudes (see Fig. 5) and an NIR capable mission would help 
greatly to uncover this difficult region. 
The Cepheids are also crucial for determining the distance scale which is critically important for dark energy studies. 
The distance scale will be anchored on Gaia parallaxes soon, but the key uncertainty remains: extinction. NIR photometry will help to 
resolve this problem and significantly contribute to a highly accurate measurement of the Hubble constant, which is a fundamental quantity for cosmology
\citep{2016ApJ...826...56R}.

Finally, AGB stars are very bright in the NIR and their study could greatly benefit from a NIR Gaia-like mission.
Mira variables show a well defined period-luminosity relation making them important galactic and extragalactic 
distance indicators. Moreover, they are much more numerous and brighter than Cepheids. So they could really 
help in the study of the Galactic structure, and extragalactic studies.
However, the period-luminosity relation for more obscure AGBs (OH/IR stars), which are even brighter than Miras 
but very extinguished at optical, are not well known. GaiaNIR will help to characterize these stars and better 
know their period-luminosity relation. So they could be used as a standard candles improving the distance scale. 

\subsection[]{The Solar System\footnotemark}\label{solarsystem}
\footnotetext{This section contains private communication from George Kaplan and Paolo Tanga mostly extracted from \cite{2014arXiv1408.3302H}.}

Astrometry has been historically one of the main tools to develop planetary science, due to the need of accurately 
computing the orbits of major and minor planetary bodies in our solar system. Among modern challenges that require 
accurate astrometric measurements, determining the masses of asteroids, stellar occultations by asteroids, the 
computation of the orbits of potentially hazardous near-Earth objects, which may exhibit very fast apparent motion, 
and, conversely, the determination of the orbits of very slow and faint bodies as the so-called Kuiper Belt Objects (KBOs) 
\citep{2007A&A...474.1015T}.
 
Ground based astrometry can achieve $\sim$~1~mas accuracy using normal telescopes although better is possible if adaptive 
optics are employed \citep{2014arXiv1408.3302H}. However, with sub-mas accuracies that can be obtained by Gaia the masses for several 
tens of asteroids can be determined, tidal effects on planetary satellites and even non-gravitational forces (Yarkovsky effect) on 
small NEOs can be studied. Effects of shape and barycentre-photocentre discrepancy can be determined if photometry and
astrometry are available \citep{2005ESASP.576..301K} thus improving the asteroid modelling and eventually the orbital accuracy

The study of orbits for over 100,000 objects in the Solar System measured by Gaia will be greatly improved if they can be based on the 
positions from two global astrometry space missions separated by 20 years. Precise fundamental ephemerides of the 
planets in the Solar System rely mostly on radar and spacecraft data for the inner Solar System, spacecraft data for Jupiter and Saturn, 
and lunar laser ranging for the Moon. The link to the ICRF is provided by VLBI observations of the spacecraft \citep{2014arXiv1408.3302H, 2015HiA....16..219F}. However, precise optical astrometry is still needed for the planets beyond Saturn as well as for most asteroids 
and Trans-Neptunian Objects (TNOs) (see chapter 8, page 305, by Standish \& Williams 2012 in \cite{2014AAS...22324720U}). 

\cite{2014arXiv1408.3302H} noted that infrequent spacecraft flybys of these objects are very useful for determining masses but do not 
constrain the orbital parameters very well. For the distant outer planets, the modern astrometric technique is to observe 
the positions of the natural satellites and infer the position of the centre-of-mass of the planet-satellite system.
Better orbits for the satellites are also of interest for modelling the dynamical history of the satellite systems.
In the inner Solar System, there is much current interest in near-Earth asteroids (NEOs) that pose some risk of 
colliding with Earth in the future. These are generally small, faint objects that do not have long-term (or any) 
observing histories, as they having orbits that are closer to the Sun than the Earth's orbit making their detection particularly challenging.
Estimating the future paths of these objects and their risk to Earth therefore requires accurate 
absolute astrometry and clearly joint missions separated by 20 years will have a very beneficial impact.

Asteroid masses can be determined by very precise astrometry before and after mutual encounters \citep{1996AJ....112.2319H}. 
However, asteroid-asteroid encounters that are useful in determining asteroid masses are infrequent events for ground-based 
observers, due to the limitations in the astrometric accuracy.  Unknown masses of asteroids contribute “noise” in the 
ephemeris of the planets in the inner Solar System \citep{2002A&A...384..322S}, although the last decade’s worth of ranging 
data from Mars spacecraft have allowed for improvements in the determination of the masses of the asteroids that significantly 
perturb that planet’s orbit \citep{2012IAUJD...7E..38K}. However, it has been shown that astrometry from space by Gaia is 
accurate enough to directly provide masses of the largest ~100 perturbers \citep{2007A&A...472.1017M}.

A new Gaia-like mission several years later, by providing accurate astrometry on a longer time span, would give access to 
additional mass determinations and also to the detection of subtle, non-gravitational effects, such as the radial drift due 
to thermal radiation \citep[Yarkovsky effect;][]{Delbo20081823}. Measuring such effects would give a major contribution to 
our understanding of asteroid families and to the mechanisms that lead to the injection of asteroids on Earth-crossing 
trajectories. Eventually, re-reducing existing astrometry from the ground by using stars having proper motions improved by 
GaiaNIR will further reduce the uncertainties  in the paths of stellar occultations by planets, asteroids, or TNOs \citep{2014DDA....4530509M}.
Such events provide accurate measurements of asteroid sizes and of vertical density profiles of TNO atmospheres.

Extending a systematic asteroid survey to the NIR immediately gives access to the asteroid mineralogy \citep{2009Icar..202..160D}. 
In fact, while the visible domain already provides a first classification, the absorption bands directly associated to mineralogy 
all appear beyond 1 micron. The spectral regions around 1 and 2 microns are the most important in this respect. As the spectral 
features are very wide, spectral resolution is not an important issue. On the other hand, the scientific reward is enormous, as 
our current knowledge of NIR asteroid spectra concerns the order of 1000 objects only. Being able to retrieve mineralogy 
information of 100,000s objects would be a major milestone with a direct impact on the models describing the formation and the 
evolution of the Solar System \citep{2014Natur.505..629D}. This is a possible argument for a spectrograph to be included in GaiaNIR
but this has not been considered further in this proposal (see Section \ref{spectra}).

\subsection{Proper motion patterns, real time cosmology and fundamental physics}\label{patterns}

\cite{2016A&A...589A..71B} recently simulated the Solar System with an orbital velocity of $\sim$220~km~s$^{-1}$ 
around the Galactic centre which results in an aberration effect of about 2.5 arcminutes \citep{2014ApJ...789..166P} 
in the direction of motion. The acceleration of the Solar System in its Galactic orbit causes this effect to change 
slowly, which results in a proper motion pattern for all objects on the sky 
\citep{1995ESASP.379...99B, 2003A&A...404..743K, 1997ApJ...485...87G, 1998RvMP...70.1393S, 2003AJ....125.1580K, 2006AJ....131.1471K, 2014MNRAS.445..845M}. 
This effect is also been referred to as secular aberration drift \citep{2010ivs..conf...60T} and is generally assumed 
to be towards the Galactic centre where most of the mass is concentrated but of course the unknown distribution of dark matter 
may affect the direction. 

\cite{2016A&A...589A..71B} also studied the instantaneous velocity of the Solar System with respect to the cosmological microwave 
background (CMB) which will cause extragalactic sources to undergo an apparent systematic proper motion. The effect is referred to as 
cosmological or parallactic proper motion \citep{1986AZh....63..845K}. This effect depends on the redshift, $z$, and fundamental 
cosmological parameters can in principle be determined from the motion. The velocity ($\vec{v}$) of the Solar System with respect to 
the observable Universe produces a dipole pattern in the CMB temperature with $\Delta T/T = v/c$. Observations of the CMB indicate 
that $v = 369 \pm 0.9$ km~s$^{-1}$ in the apex direction with Galactic longitude $l = 263.99^{\circ}$ and latitude $b = 48.26^{\circ}$ 
\citep{2009ApJS..180..225H, 2014A&A...571A..27P}. This motion should produce a parallactic shift of all extragalactic objects towards the antapex.

Both the acceleration of the Solar System and the cosmological proper motion give rise to dipole patterns in the proper motions of distant objects. 
However, the former does not depend on the redshift, while the latter does, which makes it possible to separate the effects.
\cite{2016A&A...589A..71B} showed that this proper motion effect is only just within the theoretical limits of the Gaia mission but the centroiding
of extended Galaxies may result in the effect being too difficult to measure. However, improving the proper motions by a factor of 14 with a 
new mission 20 years later would bring such measurements safely within the reach of astrometry.

\cite{1966ApJ...143..379K} and \cite{2011A&A...529A..91T} pointed out that an anisotropic expansion of the Universe would result in a distortion, 
as a function of redshift, of all distant objects on the celestial sphere in a particular direction. The effect gives a direct measurement of 
space time curvature, which is similar to a gravitational lens, but in this case is due to the cosmological curvature and not to a single body. 
The time-dependent components of the distortion would result in patterns of proper motion that could be a function of the redshifts and can be 
measured in principle. \cite{2011A&A...529A..91T} has shown that the dipole term does not vary significantly and agrees with the predicted estimates of 4--6~$\mu$as~yr$^{-1}$. For the quadrupole anisotropy, which could be interpreted as an anisotropic Hubble expansion or as an indicator of primordial 
gravitational waves, no detection has yet been made.

\cite{2009PhRvD..80f3527Q, 2012PhR...521...95Q} used the term cosmic parallax for the varying angular separation between any two distant sources, caused by the anisotropic expansion of the Universe. They considered two different scenarios:
\begin{itemize}
	\item Bianchi 1 models in which the observer is centrally embedded in an intrinsically anisotropic expansion of the early Universe. In this case, the anisotropic stress of dark energy can induce an anisotropic expansion of the Universe at late times that cannot be constrained by the CMB background measurements. Their simplified model, assuming an anisotropy of about 1\%, gives a proper motion pattern of about 0.2~$\mu$as~yr$^{-1}$.
	\item Lemaitre-Tolman-Bondi (LTB) void models in which the observer is off-centre and the Universe is inhomogeneous and isotropic. For the models considered by \cite{2012PhR...521...95Q}, they derived effects of about 0.02~$\mu$as~yr$^{-1}$. 
\end{itemize}
The anisotropic expansion of the Universe requires (sub-)$\mu$as astrometry combined with sufficient time baseline so that it remains 
beyond the reach of even a joint mission. However, a new Gaia-like mission would be able to put much stronger independent constraints on cosmic 
anisotropy (14 times better that Gaia alone).
		
Astrometric detection of quasars, i.e. solely from zero linear proper motion and parallax, unbiased by assumptions on spectra, 
might lead to the discovery of new kinds of extragalactic point sources. This issue has been studied by \cite{2015A&A...578A..91H} 
in order to see how well this can be done with the expected Gaia data and with the smaller proper motion errors from two missions.
Note that the quasars may have non-zero apparent proper motions and parallaxes (the latter because of time-dependent proper motions projected 
on the 1-year parallax pattern) due to the quasars internal structure, see for example \cite{2011A&A...526A..25T}.

\cite{2008AIPC.1082....3B} have developed a method using Gaia photometry by low-dispersion spectra and shown with simulated 
data, that it is possible to achieve a pure sample of quasars (upper limit on contamination of 1 in 40000) with a completeness 
of 65\% at magnitudes of G = 18.5, and 50\% at G = 20.0, even when quasars have a frequency of only 1 in every 
2000 objects. 

With a different approach \cite{2015A&A...578A..91H} aimed to discover astrophysically interesting quasars in the remaining sample. 
The incompleteness of quasar samples based on selection by optical photometry has been studied intensively for many years and it 
is now well established that such samples miss a substantial number of, in particular dust-reddened, quasars (see, e.g., \cite{2013ApJS..204....6F} for a recent study). 
Quasar candidates are selected solely on the basis of their lack of proper motion. This selection strategy also has the potential 
to select other extragalactic point sources, e.g. potentially new classes of objects. 

To examine the feasibility of this approach they determined the number of false positives, e.g. how many stars will be selected 
in this way and where on the sky (or towards which galactic coordinates) will the problem of stellar contamination be most severe.
They used an artificial catalogue generated for the Gaia mission (the so called GUMS data, the ``Gaia Universe Model Snapshot” \citep{2012A&A...543A.100R}).
This enabled them to derive precise numbers for the expected true detections and for the false detections due to stars which happen 
to show zero motion. \cite{2015A&A...578A..91H} predicted that two missions would be 100 times better than one Gaia mission since the 
number of false detections will be 100 times smaller. This follows because the proper motion errors were assumed to be 10 times smaller 
in both coordinates. For regions above 30 degrees latitude the ratio of quasars to apparently stationary stars is above 50\% and towards the 
poles about 80\% when using Gaia data. With a Gaia successor in 20 years the ratio would improve dramatically at all latitudes but 
crucially an NIR option would also allow us to probe the low latitude regions more completely. With GaiaNIR 14 times smaller errors 
for proper motions are predicted and consequently even better results for the quasar selection.

A number of tests for fundamental physics could be explored with a new Gaia mission and the higher proper motion accuracy could allow current Gaia tests to be repeated with higher accuracy.
A few examples are listed below:
\begin{itemize}
	\item More accurate tests in the weak field limit of the Solar System (PPN-$\gamma$ \& PPN-$\beta$, time dependence of the gravitational constant). 
	\item Cosmic expansion can be observed or constrained and alternative gravity theories could potentially be tested.
	\item Strong field gravity tests at Galactic centre in the NIR.
\end{itemize}

\section{Scientific requirements}\label{requirements}

The fundamental science  requirements of the NIR astrometry mission are based on the Gaia experience. Since the science case depends on repeating the Gaia-like astrometric performance (with an additional band-pass extension into the NIR), we start with a recap of the Gaia mission.

Gaia was successfully launched in December 2013 and has a nominal mission operational duration of 5 years.
During this time it  will continue to survey more than one billion objects relatively uniformly on the sky. Gaia  operates in the 
magnitude range from 6 to 20 and should reach end of mission accuracies of 20--25~$\mu$as at G=15 for positions and parallaxes at 
mean mission epoch. Similar accuracies for the proper motions of bright sources (down to 15th magnitude) are expected in $\mu$as~yr$^{-1}$.

Gaia measures unfiltered white light from about 350--1000 nm giving G-magnitudes and also has dedicated photometric 
and spectroscopic instruments on board. The photometric instrument uses low-dispersion prisms to yield the integrated flux of a low-resolution 
blue and red-photometers (BP, RP) in the ranges 330--680~nm and 640--1000~nm giving broad G-magnitudes ${\rm G_{BF}}$ and  ${\rm G_{RP}}$, 
respectively. In addition there is a radial velocity spectrometer operating in the wavelength range 847--874~nm and the integrated 
flux gives photometric narrow band ${\rm G_{RVS}}$ magnitudes \citep[see][for more details]{2010A&A...523A..48J}.
\begin{center} 
	\begin{figure}[h]
		\centerline{\includegraphics[scale=0.4]{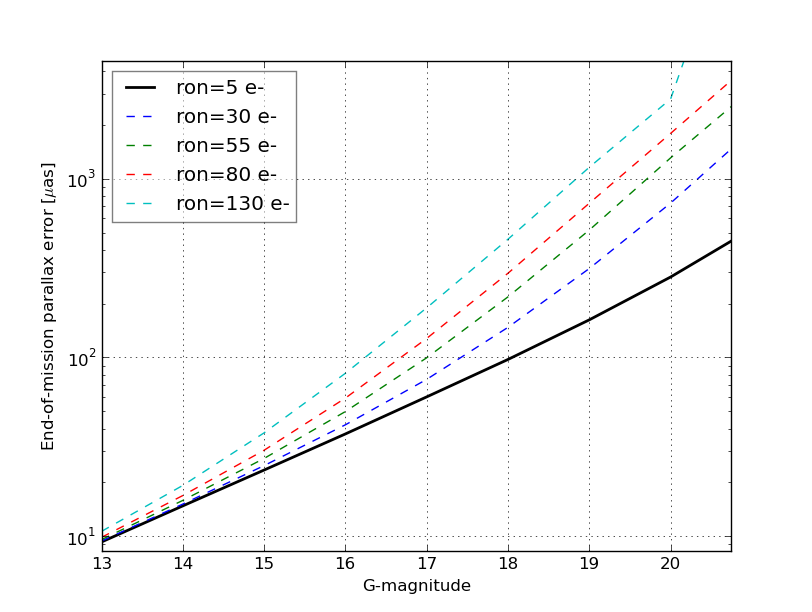}}
		\caption{\em{Standard parallax errors derived at the end of a five year Gaia-like mission. The curves were generated using  Monte Carlo simulations to estimate the the centroiding precision on a Gaia-like PSF for a range of detector readout noise levels. The thick black curve approximates that of the expected standard errors at the end of the Gaia mission. It is readily apparent that for the results of a future mission  to be comparable to those of Gaia at the faint end then a readout noise of $\lesssim10-15$~e$^{-}$~rms is required. See Section~\ref{instrument} for  detail on the practical implications of this for NIR detectors operating in TDI mode. The astrometric standard error curves presented here are derived using arguments presented in  \citet{2012Ap&SS.341...31D}.
		}} \label{Fig:ronSims}
	\end{figure}
\end{center}
	
\subsection{Requirements on the astrometric instrument}\label{astro_requirements}

The astrometric performance of the proposed mission should be comparable to or better than Gaia, with global micro-arcsecond astrometry performed for $\sim$1\% of the Galactic stellar population. Figure~\ref{Fig:ronSims} shows the standard errors on the parallax measurements expected after a five year Gaia-like mission (black curve). Of course, the final shape of this curve will depend on the scanning law used during the mission, the detector performances, payload stability, etc. However, the key point is that the astrometric requirements are similar to Gaia.  So the ability to meet these requirements in terms of spacecraft operations, onboard autonomous detection, payload stability, data-processing etc. are well demonstrated, the new challenge lies in the extension of the band pass into the NIR, where the quantum efficiency of the detectors should be greater than $85$\% up to at least 1600~nm. This minimum quantum efficiency is needed in order to meet the science requirement that a centroiding precision gain of at least 2.9 magnitudes (relative to Gaia's performance)  should be obtained for a G2V star with an extinction of A${\rm _V}=10$.   The origin for this requirement is further discussed in Section~\ref{Measurement} and  the challenges that it imposes on detector performance are discussed in more detail in Section~\ref{Detectors}. The read-out noise for each detector should reach no more than $12.5$~e$^{-}$ and the dark current and sky background (assumed to be V=22.5 mag/arcsec$^2$) should be negligible given the low temperature, and there should be good stray-light control. Passive cooling is required to keep the satellite costs low and the potential mission lifetime long (5--10 years).

\subsection{Requirements on the photometric instrument}\label{photo_requirements}

In order to meet the astrometric requirements that are described above it is necessary to obtain information on the spectral energy distribution of each object which is required to correct the systematic chromatic shifts that are induced by the optical system. Therefore, it is required that a four-band photometric instrument is also included in the design in order to provide mmag accuracy and to help disentangle the astrophysical parameters. 
For further detail on the photometric measurements see Section~\ref{Photometry}.

\subsection{Requirements on the spectroscopic instrument}\label{spectro_requirements}

Unlike Gaia, there is no requirement for the inclusion of a spectroscopic instrument, see Section~\ref{spectra} for further detail.

\section{Measurement concept}\label{Measurement}

The measurement concept for obtaining global astrometric measurements from space (which is mandatory) is well-defined and well demonstrated \citep[see, for example,][]{2001A&A...369..339P}. Indeed, the concept is identical to that of the currently flying Gaia mission where the nominal mission lifetime is 5--6 years. However, for Gaia, a proposal is currently being prepared to continue observations for a total of 10 
years if fuel consumption and the hardware on-board continue to operate as expected. Such a mission will effectively double the number 
of measurements giving an accuracy improvement by a factor of $\sqrt{2}$ in the positions and parallaxes but a factor  of $2\sqrt{2}$
in the proper motions which also benefit from a doubling of the measurement baseline. A simple calculation shows that a combination 
of two 5 year missions (labelled with subscript N for GaiaNIR and G for Gaia), 
assuming a positional and a proper motion accuracy of 25~$\mu$as~(yr$^{-1}$)\footnote{This is the end of mission (5 years) accuracy. 
The selected value does not matter as the fractional improvement is independent of the value used.} and a 20 year baseline will 
give \citep[see equation 2 in][]{2015AJ....150..141F}:
\begin{align}
&&
\sigma_{\mu_\alpha *} = \frac{\sqrt{\sigma_{\alpha^*_{\rm N}}^2 + \sigma^2_{\alpha^*_{\rm G}}}}{t_{\rm N} - t_{\rm G}}
= \frac{\sqrt{25^2 + 25^2}}{20} \sim 1.77~\mu{\rm as~yr}^{-1} \, ,
&&
\sigma_{\mu_\delta} = \frac{\sqrt{\sigma_{\delta_{\rm N}}^2 + \sigma^2_{\delta_{\rm G}}}}{t_{\rm N} - t_{\rm G}}
= \frac{\sqrt{25^2 + 25^2}}{20} \sim 1.77~\mu{\rm as~yr}^{-1} \, 
\end{align}
which is a factor of 14 better in both proper motion components. If we then assume two 10 year missions one gets
an extra factor of $\sqrt{2}$ improvement in the individual positions from each mission giving an overall improvement by a factor of 20 
for both proper motion components compared to the initial values of 25~$\mu$as~yr$^{-1}$. One of these missions is already launched and 
if a new mission follows we would get these improvements in proper motions for more than a billion stars. Parallaxes will 
also improve mainly due to the additional measurements \citep{2014arXiv1408.2190H} but also because the proper motions are much 
more accurate. The parallaxes can then be slightly better determined (an indirect improvement) which has already been demonstrated in 
the TGAS solution \citep{2014A&A...571A..85M, 2015A&A...574A.115M} for Gaia where Hipparcos/Tycho-2 data are combined with Gaia's to form 
the first data release. Systematic errors will be present in both Gaia and GaiaNIR data, but they should be uncorrelated 
and it may be possible to use a joint solution of both missions to partially reduce these systematic errors.

\begin{figure}[tbh]
	\begin{center}
		\includegraphics[scale=0.25]{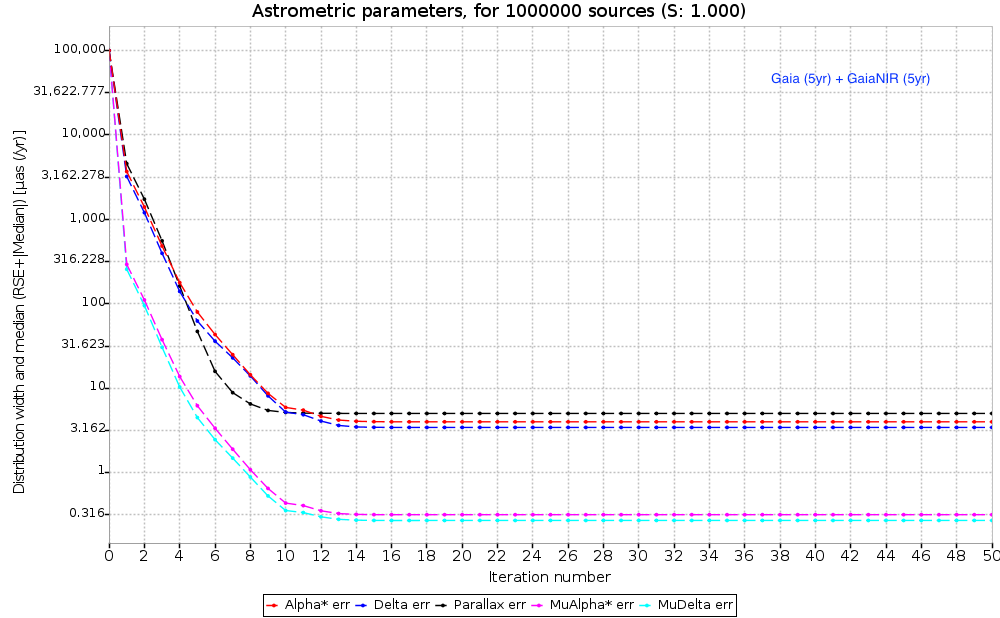}
		\includegraphics[scale=0.25]{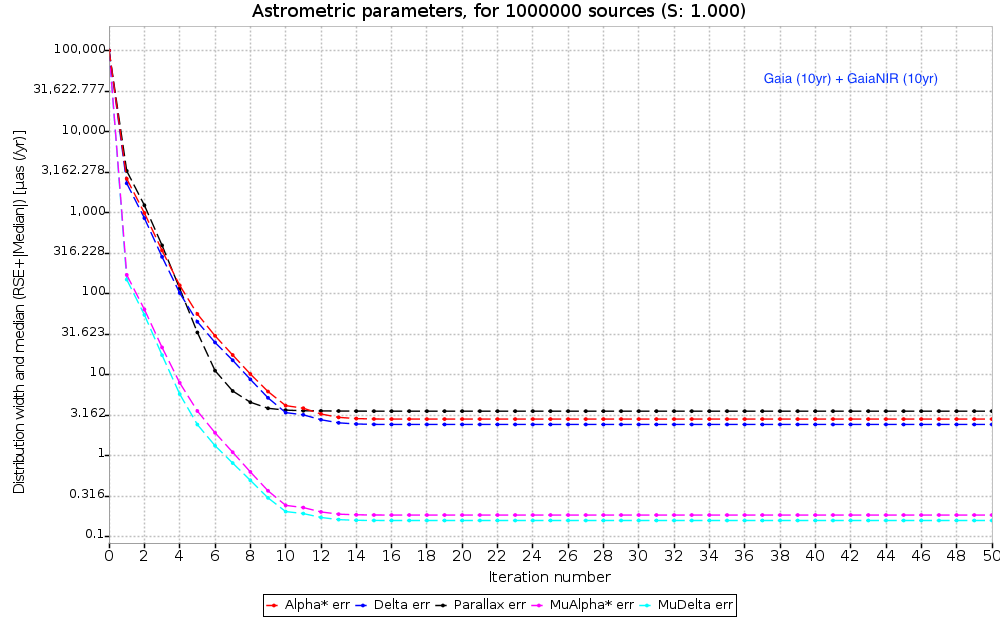}
		\caption[]{\em{Simulated astrometric results of a Gaia+GaiaNIR mission for 5+5 years (left) and 10+10 years (right). 
				The simulations include 1 million sources randomly distributed on the sky at G=13 and have regularly been used to make sensible predictions for the real Gaia mission. In the case of a 5+5 year joint mission the proper motions reach the level for ($\mu_{\alpha*}$, $\mu_\delta$, $\varpi$) of (0.328, 0.272, 5.027) $\mu$as~(yr$^{-1}$) RSE while the case of a 10+10 year joint mission gives (0.185, 0.158, 3.541) $\mu$as~(yr$^{-1}$) RSE.
				The corresponding results for a 5 year and a 10 year Gaia mission have also been computed and for ($\mu_{\alpha*}$, $\mu_\delta$, $\varpi$) are (4.08, 3.62, 7.25) $\mu$as~(yr$^{-1}$) RSE and (1.40, 1.20, 5.03) $\mu$as~(yr$^{-1}$) RSE respectively which are close to the expected factor of $2\sqrt{2}$ improvement for proper motion and a few percent better than the $\sqrt{2}$ improvement for parallax.}
		}\label{Fig:convergence}
	\end{center}
\end{figure}
Figure \ref{Fig:convergence} shows the results (expressed as RSE\footnote{The Robust Scatter Estimate (RSE) is defined as 0.390152 times the
difference between the 90th and 10th percentiles of the distribution of the variable. For a Gaussian distribution it equals the standard deviation.}) 
from AGISLab \citep{2012A&A...543A..15H} 
simulations of joint Gaia+GaiaNIR missions using periods of 5 and 10 years for 
each mission segment respectively. The results show that the improvements in the final astrometric errors for proper motion are roughly 
factors of 14 and 20 better than the a 5 year Gaia mission alone as discussed above. Likewise, the improvements in positions and parallax follow 
the prediction of $\sqrt{2}$ also mentioned above. These simulations show that the simplistic estimates given above are realistic for
real astrometric missions.

This approach brings the advantage of a largely known and already implemented mission concept \citep{2012A&A...538A..78L}, 
which can be improved based on the Gaia experience (particularly by going further into the NIR). This proposal may also fit a lower 
cost envelope but this remains to be seen, as maintaining the engineering and scientific knowledge from Gaia is not trivial.
The lessons learned from Gaia will be invaluable and improvements in the data processing and instrument modelling
can be built on an already well developed processing chain. To accomplish this goal we need to build a new Gaia-like mission 
to fly around 2035 or slightly later. There are obvious improvements that can be made in such a mission. 
\begin{enumerate}
\item Adding the possibility of making NIR measurements using TDI mode to probe the obscured regions of the Galaxy. Passive cooling of the payload is desirable to avoid excessive costs and its feasibility requires study. 
\item Improving the angular resolution of the photometry by using 4-band photometry to give all-sky coverage for a 
billion objects with a single instrument. The inclusion of a NIR band would broaden the scientific scope and the selection of the bands
should be part of this study. 
\item The addition of an on-board spectrographic instrument could be considered but is not essential. There are a large number of high quality ground 
based spectrographs currently being developed. It should be noted, however, that ground based surveys aim to collect spectroscopy 
from tens of millions of objects in selected regions of the sky but a Gaia-like mission would survey over a billion stars with 
uniform sky distribution but lower resolution using a single instrument.
\item A key question for this study is: can a similar or better astrometric accuracy (as good as a few $\mu$as in position and parallax) 
be achieved in the optical and NIR with TDI mode by 2035?
\end{enumerate}

\subsection{Photometry and Spectroscopy}\label{Photometry}
For a new Gaia-like mission it is proposed that the photometry with low-dispersion spectra could be replaced by filter photometry in 
4 bands. This will provide photometry of all sources, sufficient for the required chromatic corrections of astrometry and it 
will give photometry of narrow double stars which cannot be obtained from the ground nor can it be obtained with Gaia because the 
long spectra of the two stars overlap. Ground-based surveys of multi-colour photometry and spectra will be available for astrophysical 
studies for a large fraction of the stars but Gaia-like precision and accuracy, in the mmag realm, is much better than any ground-based 
survey. In addition, there is nothing comparable to Gaia blue and red photometry in the NIR, which could enable vast new science cases.
There are several reasons to perform photometry of sources with the mission itself:
\begin{enumerate}
\item To enable chromatic corrections of the astrometric observations.
\item To provide astrophysical information for all objects, including astrophysical 
classification for objects such as star, quasar, etc. 
\item To provide astrophysical characterisation, e.g., interstellar reddening, effective 
temperatures for stars, photometric redshifts for quasars, etc.
\end{enumerate}
\begin{figure}[tbh]
	\begin{center}
		\includegraphics[scale=0.4,trim={0cm 0.5cm 0cm 0cm},clip]{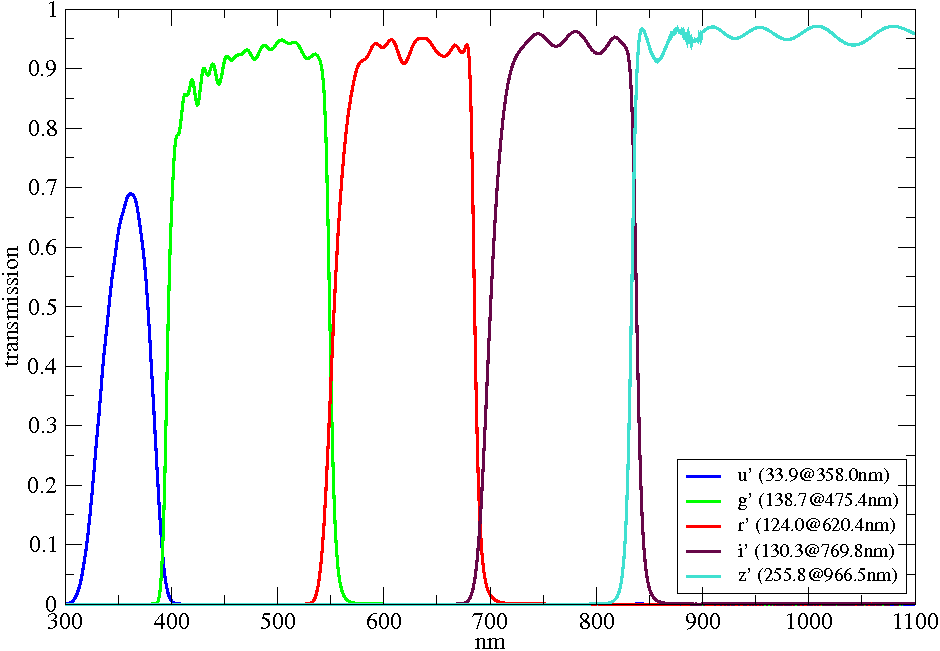}
		\caption[]{\em{Transmission of the Sloan u$'$, g$'$, r$'$, i$'$, z$'$ filter bands (from this  \href{http://www.aip.de/en/research/facilities/stella/instruments/data/sloanugriz-filter-curves}{link}).}}\label{Fig:bands}
	\end{center}
\end{figure}

For a new mission the first reason is still valid to the same extent as for Gaia, but the justification for the other reasons has changed.
By the time of a new mission, multi-colour photometry will be available for many of the observed stars with better 
spectral resolution than the mission itself can provide. Such photometry will be provided and be available from large surveys 
such as Pan-STARRS and LSST, providing five or six spectral bands from 300 to 1100 nm, but both surveys have bright saturation limits 
at 15 and 17, respectively. The angular resolution of these surveys cannot be much better than 0.5 arcsec since they are ground 
based and they do not cover the full sky. However, both Gaia and Gaia-NIR will be all-sky surveys and very importantly will be at the 
mmag accuracy level. For Gaia the angular resolution along scan of the astrometric observation is about 0.12 arcsec (FWHM of the 
sampled line-spread function). For photometry however the low-dispersion spectra limit the resolution greatly for double and 
multiple stars because the spectra of the two stars will overlap, each spectrum having a length of 1--2 arcsec. 

For a new mission, it is proposed to consider filter photometry which gives higher resolution in crowded fields 
due to far less blending. The two prisms on Gaia could be replaced by four filters as suggested by  H{\o}g (2016)\footnote{See this \href{https://dl.dropboxusercontent.com/u/49240691/GaiaNIRAccu.pdf}{link}.}. H{\o}g (2016) has pointed out that the four Sloan bands,
g$'$, r$'$, i$'$, z$'$ (see Figure \ref{Fig:bands}) or similar, would provide a realistic estimate of photometric accuracy for adjacent 
non-overlapping bands and are better adapted to the CCD sensitivity than the classical B, V, R, I bands which were designed for photo 
cathode detectors. The u$'$-band cannot be included because the silver coated mirrors will not transmit anything. These bands are narrower 
than is possible with three bands which could therefore give better accuracy per band, but the four bands are astrophysically more useful.
The 4-band photometry can be subtracted from the broad G$_{\rm NIR}$-band which would effectively provide a 5 band filter. For example, 
using the Sloan g$'$, r$'$, i$'$, z$'$ filters the J band can effectively be obtained by subtraction. The detailed choice of the filter 
bands needs to be done after the NIR study  which will decide which wavelength bands can be implemented. The suggestion for photometry 
bands above may be modified depending on a detailed trade-off between achieving accurate chromatic corrections and the possibility of 
getting good astrophysical information at the same time.

The accuracy of such filter photometry will be much better than the Gaia photometry from the short spectra in three respects:
\begin{enumerate}
\item All stars, especially the faint ones, will be less affected by noise from readout, background and parasitic stars.
\item The accuracy of filter photometry is much better for faint stars, e.g., for G2V stars at G=20 the accuracies of g$'$ 
and BP are about 28 and 56 mmag respectively. At G=19 the numbers are 13 and 23 mmag.
\item For components of double and multiple stars the improvement will be most pronounced, and they will obtain good 
photometry in many cases where Gaia could not give anything. A further advantage of the simpler data reduction is worth mentioning.
\end{enumerate}
It must be emphasized that the 4-band photometry is meant to make the chromatic astrometric corrections 
and to give excellent astrophysics. Photometry was an historical necessity, when spectra were too expensive to obtain. 
Dedicated spectrographs are now being built which will obtain millions of stellar spectra in the optical 
and IR (see Section \ref{spectra}) that will provide the quantitative science. However, even though it may still make sense to obtain 
spectra from an astrometric mission which would globally survey over one billion objects with very high angular resolution, 
this is not a driving force in the current proposal.

\subsubsection{Photometric distances}\label{photodistances}
Distances of stars are needed to derive tangential velocities from proper motions. 
Distances may be obtained from the parallaxes observed with Gaia, but they must be 
supplemented by photometric-spectroscopic distances for remote stars. The latter 
method will be much strengthened through calibration with accurate trigonometric 
distances from Gaia which will provide, e.g., better than 1.0 percent accuracy for 
10 million stars, most of these will be dwarfs. For giants of luminosity class III, 
the sample within 2 kpc from the Sun will contain stars brighter than V=11.5 if we assume 
M=0.0 mag to be typical. With $\sigma_\pi$=7.1 $\mu$as a relative accuracy better than 7.1/500
or 1.4 percent is then obtained for distances of giants. More precise estimates of the number 
of stars may be obtained from the population synthesis Galaxy star count model TRILEGAL 
\citep{2005A&A...436..895G,2012ASSP...26..165G} as described in Sect.3.1. of \cite{2014ApJ...797...14P}. 

It will be crucial for the determination of photometric distances to these stars that their 
absolute magnitudes can be accurately calibrated. The following studies of photometric calibration 
in clusters are available in \cite{2002ASPC..274..288B}, \cite{2004MNRAS.351.1204V,2004MNRAS.354..815V} and \cite{2007ApJ...661..815R}. 
More recent studies summarized in H{\o}g (2015)\footnote{See this \href{https://dl.dropboxusercontent.com/u/49240691/GaiaMag.pdf}{link}.} 
show that distances with 20\% accuracy are now obtained for normal single stars, main sequence and giants. 
An accuracy of 10\% will perhaps be common standard by 2025 after calibration with Gaia data and an accuracy of 
1 or 2\% may even be obtained for giants and some other types of stars.

The expected photometry from Gaia is presented by \cite[][in Fig. 19]{2010A&A...523A..48J} and it appears the photometric errors 
at G=19 are about 3 mmag for G and 8 mmag for the low-dispersion red photometer (GRP), but much larger for the blue band 
photometer (GBP)\footnote{The first data release of the actual photometry from Gaia is due on September 14th, 2016 and will be 
published in Evans et al., A\&A, (2016) and van Leeuwen et al., A\&A, (2016).}.
Absolute magnitudes may therefore be obtained from the spectra of faint red stars \cite{2014arXiv1408.2190H}.
The main source for accurate absolute magnitudes of faint Gaia stars, after luminosity calibration with the Gaia parallaxes, 
would probably be the many photometric and spectroscopic surveys. Spectrograph examples include 
\citep[4MOST,][]{2012SPIE.8446E..0TD}, 
\citep[WEAVE,][]{2014SPIE.9147E..0LD}, 
\citep[MOONS,][]{2014SPIE.9147E..0NC}, 
\citep[APOGEE I-II,][]{2015arXiv150905420M}, etc.

\subsubsection{Spectroscopic radial velocities}\label{spectra}

Spectra for radial velocities plus photometry for classification and distances of faint stars may be obtained in great number 
with the ESOs wide-field, high-multiplex (1500--3000 fibres) spectrograph 4MOST \citep[see Fig. 1][]{2012SPIE.8446E..0TD} 
which will be running nearly full time on the 4m-class telescope (VISTA) from about 2020.  Radial velocities are required 
to obtain the full 3-D space velocities for unambiguous dynamical studies and accuracies of $\le$~2~km~s$^{-1}$ for the 
faintest stars observed by Gaia may be obtained with 4MOST. 
WEAVE \citep{2014SPIE.9147E..0LD} is a new wide-field multi-object spectrograph for the William Herschel Telescope and will 
do optical ground-based follow up of the LOFAR and Gaia surveys. WEAVE will host 1000 multi-object fibres which are fed to a 
single spectrograph, with a pair of 8k (spectral) x 6k (spatial) pixel cameras. The project is now being developed with 
commissioning of the telescope expected in 2017. Unfortunately, 4MOST and WEAVE will only operate in the wavelength 
region 390--1000~nm and 370--1000~nm respectively and will not probe the NIR to open new science cases in the Galactic centre.

For NIR spectrographs, Multi-Object Optical and NIR Spectrograph (MOONS) \citep{2014SPIE.9147E..0NC} on the Very Large Telescope (VLT) 8.2 meter 
telescope will operate in the wavelength range: 800--1800~nm. The large collecting area offered by the VLT, combined with the 
large multiplex and wavelength coverage will constitute a powerful, unique instrument able to pioneer a wide range of Galactic, Extragalactic 
and Cosmological studies and provide crucial follow-up for major facilities such as Gaia.
APOGEE-I \& II \citep{2015arXiv150905420M} operates from 1510--1700~nm using high-resolution, high signal-to-noise IR spectroscopy to
penetrate the dust that obscures significant fractions of the disk and bulge of our Galaxy to survey over 100,000 red giant stars across 
the full range of the Galactic bulge, bar, disk, and halo. 

The combination of these instruments would provide the third dimension of the space velocities needed particularly in the dusty regions
that currently cannot be explored by Gaia and at first sight it would seem unnecessary to include a spectrograph in a GaiaNIR design.
However, a GaiaNIR design could provide spectra for up to one billion objects in the optical and NIR globally on the sky, albeit at low resolution, 
which would provide this third component of the space velocity for most observed objects compared to some tens of millions of spectra from selected 
areas with different instruments. GaiaNIR would potentially provide a unique opportunity to obtain photometry and spectroscopy   
with a single instrument making the calibration of the science results uniform across the sky. However, we have concluded that the spectroscopic
option for GaiaNIR is a nice addition (presumably at considerable cost) but is not a necessary requirement to achieve the science goals outlined 
in this document.

\subsection{Potential instrument design and sensors}\label{instrument}

H{\o}g et al. (2015\footnote{See this \href{https://dl.dropboxusercontent.com/u/49240691/GaiaNIR.pdf}{link}.} and 2016\footnote{See this \href{https://dl.dropboxusercontent.com/u/49240691/GaiaNIRAccu.pdf}{link}.}) have considered potential 
focal plane assemblies that could be used for a Gaia successor mission with all sensors covering 400 nm~to NIR about 2000~nm. 
In their papers they considered a number of possible configurations both with combined optical and NIR detectors and 
separate sensors (CCD and HgCdTe-CMOS) for the two main wavebands. 

\begin{center}
	\begin{figure}[tbh]
		\includegraphics[scale=0.2377]{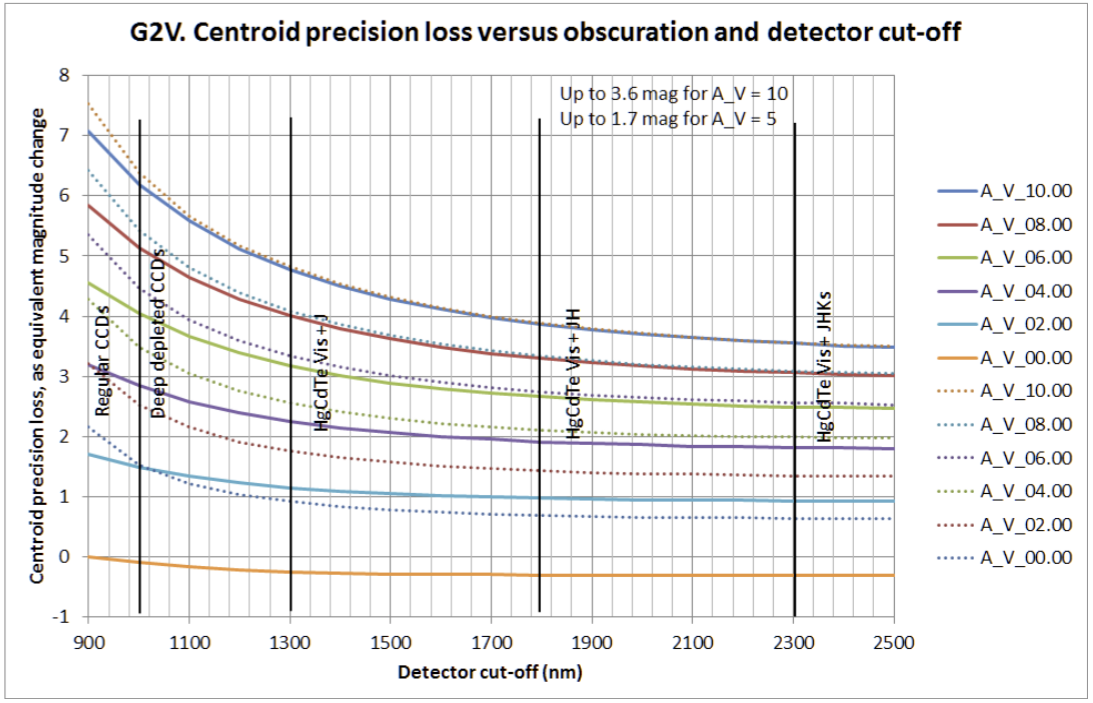}
		\includegraphics[scale=0.23]{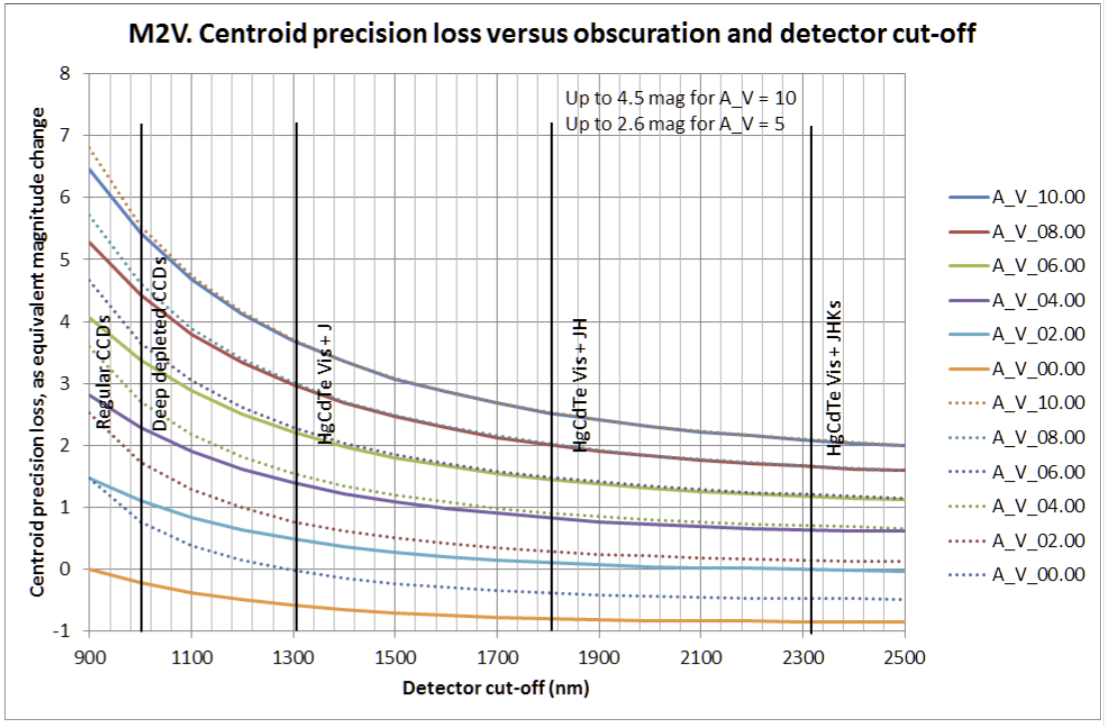}
		\caption[]{\em{The left figure shows the precision of a G2V star while the right figure shows the same for an M2V star.
				The solid line shows values for optical plus NIR (starting from 400~nm) while the dotted line is for NIR only 
				starting from 800~nm 
				(A. Mora, Session 5, 2015)\footnotemark.}}\label{Fig:accuracyStars}
	\end{figure}
\end{center}
\footnotetext{See this \href{http://great.ast.cam.ac.uk/Greatwiki/GaiaScienceMeetings/FutureAstrometryJul15?action=AttachFile&do=get&target=astrofutures-jul15-mora.pdf}{link}.}

They arrived at a focal plane assembly consisting of HgCdTe detectors with the band gap tuned in order to provide high quantum 
efficiencies. The preliminary design placed these detectors in the whole astrometric field and were designed to 
detect all wavelengths from 400 to about 2500~nm. Extension of the sensitivity towards red means that the total number of detected 
photons will be increased for all stars, the more so for redder spectral types. This effect is seen best in the lower (orange) solid curves 
compare to the lower (pale blue) dotted curves in Fig.~\ref{Fig:accuracyStars} showing an improved astrometric precision for stars, with no extinction,
for both spectral types. The improvement becomes less as extinction increasing as fewer photons are collected in optical bands. 
The resulting photometric band for the astrometric field may be called G$_{\rm NIR}$ and this band will have a longer effective 
wavelength than the present Gaia G-band, the more so for redder spectral types. 

\begin{figure}[tbh]
	\begin{center}
		\includegraphics[scale=0.65, trim={1cm 0.5cm 0 1.0cm},clip]{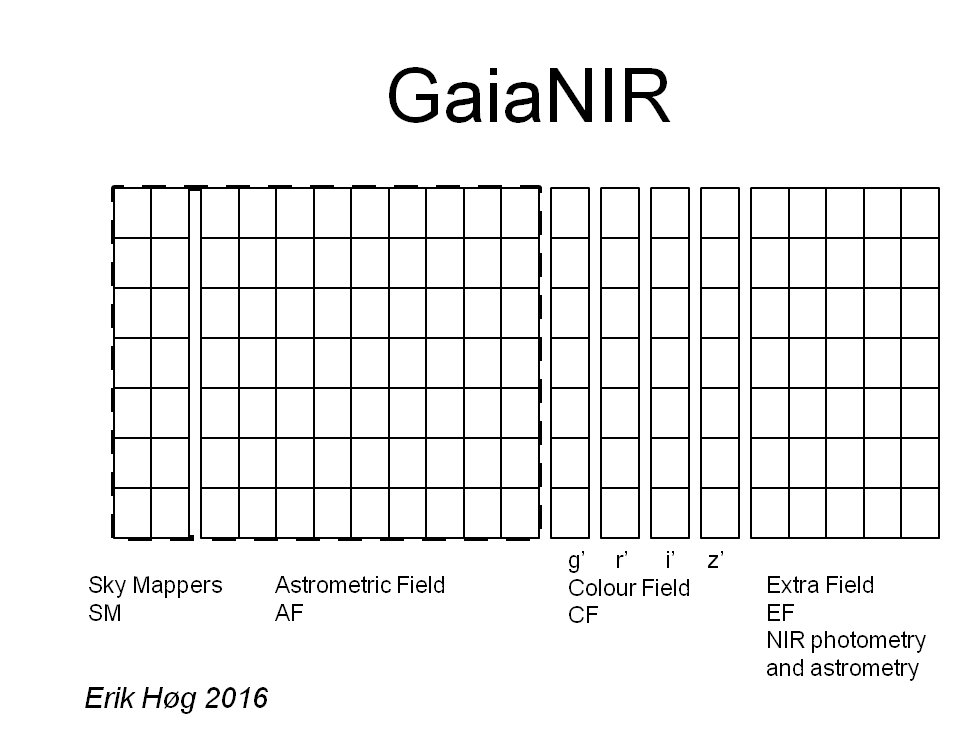}
		\caption[]{\em{Potential focal plane arrangement of GaiaNIR. Depending on the performance of the NIR sensors one could envisage a focal plane with NIR sensors in all fields. NIR photometry could also be obtained on some of the sensors in the Extra Field which would also provide more astrometric measurements. However, if the NIR sensors are problematic in terms of the read out noise achievable one could use CCDs for the astrometry and the 4 band photometry followed by NIR sensors in the Extra Field operating in a narrower band, e.g. 900--1600~nm. In that case, one of the NIR detectors must act as a SM to detect the faint brown dwarfs, very reddened stars, cool white dwarfs, etc.} }\label{Fig:GaiaNIRfa}
	\end{center}
\end{figure}

An ideal arrangement for the focal plane for GaiaNIR (see Figure \ref{Fig:GaiaNIRfa}) could include sensors operating from 400--2000~nm in all 
fields but this could be relaxed to a sensitivity from 400--1600~nm in all fields while still maintaining the science goals.  As the performance 
of the NIR sensors is potentially problematic due to readout noise, optical CCDs could be used as a fall back option for the astrometry and 
for the g$'$, r$'$, i$'$, z$'$ photometry with sensitivities similar to that on Gaia. In that case NIR sensors sensitive from 900--1600~nm 
could be used in a dedicated Extra Field providing NIR astrometric and photometric measurements. Note that in that case, one of the NIR 
detectors must act as a SM to detect the faint brown dwarfs, very reddened stars, cool white dwarfs, etc.

The astrometric performance expected from a new mission with a focal plane assembly and NIR sensitivity for three different types 
of detectors is tabulated in Table \ref{Tab:centroidloss} which shows the results for regular CCDs, HgCdTe Vis+J band sensors, and 
HgCdTe Vis+HJK band sensors. The sensor types correspond to wavelengths bands of 400--900, 400--1300, and 400--2300 nm respectively.
The performance for various extinctions levels are given in the table for both G2V and M2V type stars. The centroid precision loss L, 
expressed as the equivalent magnitude change is followed by the G magnitude at which a parallax error $\sigma_\pi$=107$~\mu$as is obtained.  
The value $\sigma_\pi$=107$~\mu$as corresponds to the end of mission accuracy for G=18 in the nominal Gaia mission using regular 
CCDs and with A${\rm _V}$=0.

\begin{table*}[htbp!]
	\centering \caption[]{\em{The centroid precision loss L, expressed as the equivalent magnitude change together with the G magnitude at which a parallax error ($\sigma_\pi$) of 107$\mu$as is obtained. Taken from this \href{https://dl.dropboxusercontent.com/u/49240691/GaiaNIR.pdf}{link}.}\label{Tab:centroidloss}}
	\begin{tabular}{ccccc}
		\toprule\toprule
		Type& Ext.        &  Regular CCDs  & HgCdTe Vis+J  & HgCdTe Vis+JHKs \\
		\midrule
			& A${\rm _V}$ &  L, G          & L, G          & L, G            \\
		\midrule
		G2V & 0           &  0.0, 18.0     & -0.2, 18.2    & -0.3, 18.3      \\
		    & 5           &  3.7, 14.3     &  2.7, 15.3    &  2.2, 15.8      \\
		    & 10          &  7.0, 11.0     &  4.7, 13.3    &  3.6, 14.4      \\
		\midrule
		M2V & 0           &  0.0, 18.0     & -0.6, 18.6    & -0.8, 18.8      \\
		    & 5           &  3.4, 14.6     &  1.8, 16.2    &  0.9, 17.1      \\
		    & 10          &  6.4, 11.6     &  3.7, 14.3    &  2.1, 15.9      \\
		\bottomrule
	\end{tabular}
\end{table*}

The table shows for an extinction of A${\rm _V}=10$ mag that an astrometric precision of 3.4 and 4.3 Gmag fainter could be obtained with the NIR, 
using HgCdTe Vis+JHKs compared to regular CCDs, for stars of types G2V and M2V respectively. Furthermore, measurement of un-obscured 
M2V stars will go 0.8 mag fainter for a given astrometric precision. It is noted that different noise levels may be expected 
especially for faint objects due to diffraction, scattered light and read out noise, but this has not been taken into account in this table. 
The astrometric performance expected here assumes LSF integration, e.g. by TDI as in Gaia with pixel width along scan matching the 
shortest wavelengths. The on-board processing strategy for GaiaNIR could be very similar to that of Gaia, which has proven to work 
very well in practice, i.e. autonomous operations, detecting, measuring and downlinking all objects which are sufficiently bright.
Gating mechanisms may also need to be considered to ensure bright stars are also detected. Ideally, we would like to detect objects
from 6-21~G, as in Gaia, but much depends on the characteristics of the sensors and this will have to be part of the feasibility study.
In conclusion the study and development of a suitable NIR sensor for TDI mode should be pursued as soon as possible 
to be available for the development of such a mission.

\subsection{Photometric errors for CCDs with Sloan-like photometry}\label{PhotoErrors}

\begin{figure}[!ht]
 	\centering
 	\includegraphics[scale=0.35, trim={0cm 0.9cm 0cm 0cm}]{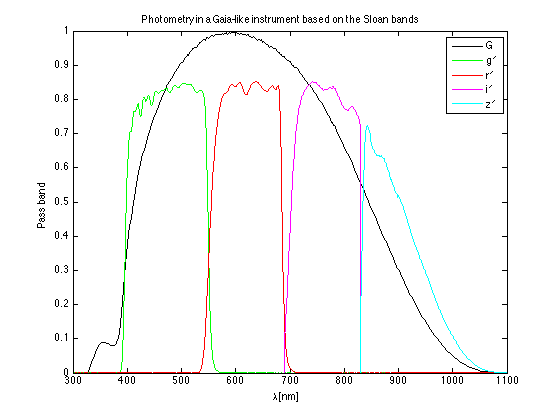}
 	\qquad
 	\begin{tabular}[b]{ccccccccc}\hline
 	G [mag]  & G   & G$^\prime$&  BP      &  RP     & g$^\prime$ & r$^\prime$ & i$^\prime$ & z$^\prime$ \\ \hline
 		3-13 & 0.2 & 0.2       &  0.8     &  0.8    &  0.6       &  0.6       &  0.6       &  0.6       \\
		14	 & 0.2 & 0.2       &  0.9     &  0.9    &  0.7       &  0.7       &  0.8       &  0.9       \\
		15	 & 0.2 & 0.2       &  1.2     &  1.1    &  1.1       &  1.1       &  1.2       &  1.5       \\
		16	 & 0.4 & 0.3       &  2.0     &  1.8    &  2.0       &  1.9       &  2.1       &  2.8       \\
		17	 & 0.6 & 0.5       &  4.1     &  3.6    &  3.5       &  3.4       &  3.8       &  4.9       \\
		18	 & 0.9 & 0.9       &  9.4     &  8.1    &  6.9       &  6.5       &  7.4       & 10.2       \\
		19	 & 1.8 & 1.6       & 22.7     & 19.5    & 13.5       & 12.9       & 14.4       & 22.3       \\
		20	 & 3.7 & 3.3       & 56.1     & 48.1    & 30.8       & 29.2       & 33.0       & 52.4$^*$   \\
		20.5 & 5.6 & 4.8       & 88.6     & 75.9    & 47.7       & 45.2       & 51.3$^*$   & 70.7$^*$   \\
		21	 & 8.6 & 7.2       &140.1     &120.0    & 66.0$^*$   & 63.5$^*$   & 69.6$^*$   & 89.0$^*$   \\ \hline
		\vspace{1mm}
 	\end{tabular}
  	\captionlistentry[table]{}
 	\captionsetup{labelformat=andtable}
 	\caption{Simulated photometric errors in mmag for a G2V star. The data for the Gaia G, BP and RP bands are the median end-of-mission standard errors 
 		and have been taken from the \href{http://www.cosmos.esa.int/web/gaia/science-performance}{ESA website}, extended to G=22.0.
 		The calculations here use G$^\prime$ which gives the error if there were four times less stray light as there will be in a Sloan band. 
 		The errors for g$^\prime$, r$^\prime$, i$^\prime$,  z$^\prime$ are then calculated using simulated Sloan bands as they would look in a 
 		Gaia-like photometric system (see figure) assuming 85\% transmission. One can see that the Sloan-like filter system is similar to the 
 		BP-RP bands at the bright end but is more accurate at fainter magnitudes. Values marked with $^*$ mean they 
 		were calculated with linear extrapolation as G$^\prime$+D was greater that the available magnitude, i.e. $\le$22.0. \label{FigTab:photoError}}
\end{figure}

In order to estimate the photometric errors using CCDs and Sloan-like pass-bands we have done the following.
The digital Sloan pass-bands have a maximum transmission of 0.95 which is surprisingly high for any real colour filter. They are
renormalized and multiplied by a more realistic maximum of 0.85 which is about the same as the RP filter in Gaia.
The four Sloan bands are approximately located at: g$^\prime$ = 390-550 nm, r$^\prime$ = 550-690 nm, i$^\prime$ = 690-830 nm, 
z$^\prime$ > 830 nm, the red edge being defined by the CCD response. The digital i$^\prime$ and z$^\prime$ filters are multiplied 
with the normalized RP pass-band to create the red edge of z$^\prime$ correctly taking into account the transmissions of the 
instrument and CCDs and to make a smooth transition from i$^\prime$ to z$^\prime$. The G band stays normalized because there is 
no filter so that it will appear higher than the Sloan bands as seen in Fig \ref{FigTab:photoError}.

The number of photons is then computed for the 5 pass-bands (G + 4 Sloan bands) for a G2V star.
These un-normalised number of photons, N${\rm _G}$ for G and N${\rm _S}$ for the Sloan bands, are used to compute 
4 ratios F=N${\rm _G}$/N${\rm _S}$, the number of photons through G divided by the number of photons through the Sloan band.
If the number of photons through the filter is F times smaller than in G; then this number is the same as the number of photons 
in the G band of a star D magnitude fainter star, where D is
\begin{align}
&&
D=2.5 {\rm log} F \, .
\end{align}
The standard errors for G given in Table \ref{FigTab:photoError} are then used to derive the errors of Sloan photometry as a function of G.
Let $\sigma{\rm _S}$(G) denote the standard error in the S band for a star of G mag. The errors $\sigma{\rm _G}$(G) are obtained from column 
2 of Table \ref{FigTab:photoError}. The error of a Sloan magnitude S$^\prime$ is then obtained as 
$\sigma{\rm _S}$(G)=3$\times\sigma{\rm _G}$(G+D) for the available G+D$\le$22.0 mag. Linear extrapolation can be used to obtain numbers outside this 
range and they are indicated by $^*$ in the table. The factor 3 arises as we assume all stars are measured by 9 CCDs in the astrometric field 
for G but using only one CCD in each colour filter. This implies $\sqrt{9}=3$ times larger errors as a function of magnitude from any filter 
if it passes the same number of photons as the G band. The read-out-noise in the filter is assumed to be the same as in G. The difference in sky 
background is perhaps small and is neglected due to low temperature and there is good stray-light control, but these assumption are being investigated.

This table, with the predicted accuracy of Sloan photometry at all G magnitudes, makes it possible to make a realistic comparison of the 
performance of Gaia prism photometry with the expected griz photometry from Sloan like bands both with respect to astrometric chromaticity 
corrections and with respect to characterization of stars. Initial estimates used the median end-of-mission photometric standard errors (labelled G)
but the stray light and sky background through a Sloan filter will be about four times smaller than through the G band (labelled G$^\prime$). 
The errors in the Sloan bands are then calculated, using G$^\prime$ instead of G, and the results show a quite significant improvement at the 
faint end of about $\sim$17\% compared to using the median stray light. The software can be modified to compute the tables for NIR sensors 
instead of CCDs when reasonable assumptions become available. The z$^\prime$ band can then naturally be expanded towards red and a further NIR 
band can be added.

\subsection{Detector performance challenges}\label{Detectors}

The quality of photo-detectors with combined optical/NIR sensitivities has seen remarkable improvement over the past decade, including their
ability to withstand radiation damage. 
The current leaders in the field are the HgCdTe (also known as MCT) detector chips produced, for example, by the American company Teledyne. 
These devices, known as hybrids, combine  the advantages provided by using a HgCdTe detecting layer with the high performances 
achieved by CMOS technology in what is known as a hybrid detector. Each pixel of the detector array is indium-bonded to its 
own silicon readout integrated circuit where the charge is converted to a voltage before the information is transferred to the 
edge of the array.   One big advantage of HgCdTe as a detector material is that the (direct) band-gap can be tuned by more 
than an order of magnitude (from $\sim0.1-1.5$~eV) by varying the ratio of Hg to Cd, meaning that the wavelength range to 
which the device is sensitive to can be optimised based on the science objectives. For further detail on these devices and 
on HgCdTe as a detecting layer see \citet{2008SPIE.7021E..0HB} and references within. 

These hybrid devices will be used on-board JWST and the Euclid mission (the NIR NISP instrument) as well as many other 
ground and space-based astronomical instruments.   High  dynamic range performance means that, even for a wide range of stellar brightnesses 
a readout noise of $\sim10$e$^{-}$ is achievable for these devices. A comparison of this readout noise value with the curves 
in Figure~\ref{Fig:ronSims} show that this is already low enough for astrometric performances to be comparable to those of Gaia. 
However, due to the fact that the current implementation of these type of devices involves readout from each pixel, then the most 
obvious implementation of a TDI functionality for these devices (off-chip summation of the signal from each pixel {\em{after}} 
each pixel readout) would result in a large increase in the readout noise for each detector transit.
Assuming $n=2048$  pixels  per TDI detector transit, each with a  readout noise, $r$, of  $10$~e$^{-}$, the   total readout 
noise would be $r \times \sqrt{n} \sim 450$~e$^{-}$, which according to  Figure~\ref{Fig:ronSims}  is much too high to achieve 
the science requirements. Gaia has $4500$ pixels per TDI integration along each CCD column, so using this number of transfers 
would result in even higher noise. Even assuming an unrealistically optimistic readout noise of $1$~e$^{-}$ and $2048$ pixels, 
the total readout noise would still be  too high at $\sim 45$~e$^{-}$, resulting in parallax errors for magnitude $20$ stars of 
more than $1$~mas.  It therefore appears that, unlike Earth observation missions where low signal TDI measurements are not a 
priority, off-chip TDI summation appears not to be a realistic proposition for this type of mission.

However, one option that could be considered would be a form of on-chip charge summation where the detecting layer (HgCdTe) is 
connected to a standard silicon-type CCD chip device that is used primarily for  charge storage and transfer, as opposed to charge 
generation. In this manner, the charge generated in the HgCdTe layer of each pixel could be transferred to the silicon pixel for 
storage.  After this happens it  is then  transferred along to the next pixel where the charge that is  generated in the HgCdTe 
layer attached to this pixel is then added to the charge packet. In this manner charge is accumulated as the source transits the 
device and the readout only occurs once at the end of pixel transfers resulting in a dramatically  reduced level of readout noise 
as compared to the off-chip TDI case. Optical plus NIR HgCdTe detecting layers have already  been incorporated in CCD-like structures, for example see \citet{2005SPIE.5726..132S}. 

Although this technology seems plausible in principle, to the best of our knowledge the science drivers have not been there in 
order for this possibility to be investigated in detail. In recent years, both in terms of hybrid NIR detectors and TDI using 
CMOS technology, there has been much progress made in Europe. Indeed, since 2010 ESA has been running a multi-activity program 
to develop a large format ($\sim2\rm{k} \times 2\rm{k}$) European NIR HgCdTe high performance device that can be used for 
astronomical applications. As a result of this  there have been rapid improvements made by a handful of companies, see, for 
example, \citet[][]{2015SPIE.9639E..0ON, 2015SPIE.9639E..0XC} as well as a number of papers soon to be published in the proceedings 
of Volume 9915 of the High Energy, Optical, and Infrared Detectors section of the SPIE series for Astronomy VII (see
\href{https://spie.org/Publications/Proceedings/Volume/9915?&origin_id=x4325&start_volume_number=9900&end_volume_number=9999}{this link} for the abstracts).
In terms of TDI functionality in CMOS technology, there is much ongoing activity in Europe, including combining CMOS and CCD technology 
in order to do on-chip low-noise TDI charge summation. A list of talks from the 2015 Nov. 4th Toulouse CMOS image sensors "Workshop on 
CMOS Image Sensors for High Performance Applications" is available
\href{http://image-sensors-world.blogspot.se/2015/10/workshop-on-cmos-image-sensors-for-high.html}{here} and an e2V technical paper describing  developments in incorporating TDI funtionality in CMOS devices is available
\href{http://www.e2v.com/content/uploads/2015/06/MAYER_IISW2015_final.pdf}{here}.
Therefore, we feel that ESA has the expertise and knowledge to advise us on the best detector options for this proposed future mission.

It is worth noting that other options for achieving low-noise TDI in the optical/NIR (other detecting materials, forms of readout etc.) 
could also be examined and the best case results compared with the  option of combined CCD/NIR detectors in the focal plane in order 
to know how to best proceed with a future mission proposal.

\section{Other information}\label{other}

None.

\newpage
\section{Contact information}\label{contact}

Principal  (PI) : 
{Dr. David Hobbs, \\ Lund Observatory, Box 43, SE-221 00 Lund, Sweden. \\ Email: david@astro.lu.se.~~~Tel.: +46-46-22\,21573}\\

The Principal Investigator will be available at the 25\% FTE level to support the study activities. Additionally, there is 
a large community of scientists, many within DPAC, which support this proposal and can be called upon to give expert advice on the
science and technical requirements.

\newpage

\bibliography{GaiaNIR}
\bibliographystyle{aa}

\newpage
\appendix
\section{Proposal writers}\label{Sec:AppA}

\begin{longtable}[l]{ll}
	Hobbs,\ D.                 & Lund Observatory, Sweden. \\
	H{\o}g,\ E.                & Copenhagen University (Retired), Denmark. \\
	Mora,\ A.                  & Aurora Technology B.V., Spain. \\
	Crowley,\ C.               & HE Space Operations B.V., Spain. \\
	McMillan,\ P.J.            & Lund Observatory, Sweden. \\
	Ranalli,\ P.               & Lund Observatory, Sweden. \\	
	Heiter,\ U.                & Uppsala University, Sweden. \\ 
	Jordi,\ C.                 & Institut de Ci\`encies del Cosmos, ICCUB-IEEC, Spain. \\
	Hambly,\ N.                & University of Edinburgh, UK. \\
	Church,\ R.                & Lund Observatory, Sweden. \\ 
	Brown,\ A.                 & Leiden Observatory, Holland.\\
 	Tanga,\ P.                 & Observatoire de la C\^ote d’Azur, France. \\
    Chemin,\ L.                & Research Fellow - INPE/MCT, Brazil. \\
	Portell,\ J.               & Institut de Ci\`encies del Cosmos, ICCUB-IEEC, Spain. \\
	Jim\'enez-Esteban,\ F.     & CAB (INTA-CSIC), Spain \\ 
	Klioner,\ S.               & Lohrmann Observatory, Germany.\\
	Mignard,\ F.               & Observatoire de la C\^{o}te d'Azur, France.\\		
 	Fynbo,\ J.                 & Niels Bohr Institute, Denmark. \\
 	Wyrzykowski,\ L.           & Warsaw University Astronomical Observatory, Poland.\\
 	Rybicki,\ K.               & Warsaw University Astronomical Observatory, Poland. \\
 	Anderson,\ R.I.            & Johns Hopkins University, USA \\ 
	Cellino,\ A.               & INAF - Osservatorio Astrofisico di Torino, Italy. \\ 
	\\
\end{longtable}

\section{Other supporting scientists}\label{Sec:AppB}
\begin{longtable}[l]{ll}
Abbas,\ U.                      & INAF - Osservatorio Astrofisico di Torino, Italy. \\
Adibekyan,\ V.                  &Instituto de Astrof\'isica e Ci\^encias do Espa\c{c}o, Universidade do Porto, Portugal. \\
Altavilla, \ G.                 & INAF-Astronomical Observatory of Bologna, Italy. \\ 
Anguiano,\ B.                   & Max Planck Institut für Sonnensystemforschung, Göttingen, Germany.\\
Arenou,\ F.                     & Paris Observatory, France. \\
Athanassoula,\ E.               & Laboratoire d'Astrophysique de Marseille, France. \\
Babusiaux,\ C.                  & Observatoire de Paris, France. \\
Bachchan,\ R.K.                 & Lund Observatory, Sweden. \\
Barrado,\ D.                    & CAB (INTA-CSIC), Spain. \\ 
Bastian,\ U.                    & ARI Heidelberg, Germany. \\
Bedding,\ T.                    & University of Sydney, Australia. \\
Bellazzini,\ M.                 & INAF - Bologna Observatory, Italy. \\ 
Bensby,\ T.                     & Lund Observatory, Sweden. \\
Biermann,\ M.                   & Zentrum f\"ur Astronomie der Universit\"at Heidelberg, ARI, Germany.\\
Blagorodnova,\ N.               & California Institute of Technology, USA. \\ 
Blanco-Cuaresma,\ S.            & University of Geneva, Switzerland. \\
Bland-Hawthorn,\ J.             & University of Sydney, Australia. \\
Blomme,\ R.                     & Royal Observatory of Belgium, Belgium. \\ 
Bombrun,\ A.                    & HE Space Operations B.V., Spain. \\
Bosma,\ A.                      & Laboratoire d'Astrophysique de Marseille, France. \\ 
Bouy,\ H.                       & University of Bordeaux, France. \\
Bragaglia,\ A.                  & INAF-OA Bologna, Italy. \\
Bruzual,\ G.                    & IRyA, UNAM, Campus Morelia, Mexico. \\
Busonero,\ D.                   & INAF - Osservatorio Astrofisico di Torino, Italy.\\
Busso,\ G.                      & Institute of Astronomy, University of Cambridge, U.K. \\
Butkevich,\ A.                  & Previously at Lohrmann Observatory, Germany.\\
Caballero,\ J.A.                & Centro de Astrobiolog\'{\i}a, Spain. \\ 
Cacciari,\ C.                   & INAF, Bologna Observatory, Italy. \\
Carballo,\ R.                   & Universidad de Cantabria, Spain. \\
Carrasco,\ J.M.                 & Institut de Ci\`encies del Cosmos - Universitat de Barcelona, Spain. \\
Carry,\ B.                      & Observatoire de la C{\^o}te d'Azur, France. \\
Casagrande,\ L.                 & Australian National University, Australia. \\ 
Casewell, S.L.                  & University of Leicester, UK. \\
Casey,\ A.R.                    & IoA, University of Cambridge, UK. \\
Clementini,\ G.                 & INAF Osservatorio Astronomico Bologna, Italy. \\
Comer\'on,\ F.                  & European Southern Observatory, Germany. \\ 
Correia,\ A.C.M.                & University of Aveiro, Portugal. \\
Creevey,\ O.                    & Observatoire de la C\^ote d'Azur, France.\\
Crosta,\ M.                     & INAF - Osservatorio Astrofisico di Torino, Italy. \\
Dafonte,\ C.                    & Universidade da A Coru\~na, A Coru\~na, Spain. \\
De Angeli\, F.                  & Institute of Astronomy, University of Cambridge, UK. \\
de Jong,\ R.S.                  & Leibniz Institute for Astrophysics Potsdam (AIP), Germany. \\
de Laverny,\ P.                 & Observatoire de la C\^ote d'Azur, Nice, France.\\
de Martino,\ D.                 & INAF Capodimonte Astronomical Observatory, Italy. \\
De Ridder,\ J.                  & KU Leuven, Belgium. \\
De Silva,\ G.                   & University of Sydney, Australia. \\
de Val-Borro,\ M.               & Princeton University, USA. \\
Debattista,\ V.P.               & University of Central Lancashire, UK. \\ 
Delbo,\ M.                      & Observatoire de la Cote d’Azur, France. \\
Drew,\ J.E.                     & University of Hertfordshire, UK. \\ 
Drimmel,\ R.                    & INAF - Oss. Astrofisica di Torino, Italy. \\ 
Ducourant,\ C.                  & LAB, Bordeaux University, France. \\
Evans,\ D.W.                    & IoA, Cambridge, UK. \\
Famaey,\ B.                     & Observatoire astronomique de Strasbourg, France. \\
Fern\'andez-Trincado,\ J.G.     & Institut UTINAM,  OSU THETA, Besan\c{c}on, France. \\ 
Figueras,\ F.                   & Institut de Ci\`encies del Cosmos, ICCUB-IEEC, Spain. \\
Font,\ A.S.                     & Liverpool John Moores University, UK. \\
Frey,\ S.                       & F\"OMI Satellite Geodetic Observatory, Hungary. \\ 
Gallart,\ C.                    & Instituto de Astrofisica de Canarias, Spain. \\ 
Galluccio,\ L.                  & Observatoire de la Côte d'Azur, France. \\
Garofalo,\ A.                   & INAF-Osservatorio Astronomico di Bologna, Italy.\\
Garzón,\ F.                     & Instituto de Astrofísica de Canarias, Spain. \\ 
Gilmore,\ G.                    & IoA Cambridge, UK. \\ 
Grebel,\ E.K.                   & Heidelberg University, Germany. \\ 
Groenewegen,\ M.A.T.            & Koninklijke Sterrenwacht van Belgi\"e, Belgium. \\ 
Guy,\ L.P.                      & Department of Astronomy, University of Geneva, Switzerland. \\ 
Halbwachs,\ J.L.                & Strasbourg Astronomical Observatory, France. \\ 
Harrison,\ D.L.                 & IoA, University of Cambridge, UK. \\
Hestroffer,\ D.                 & IMCCE/Paris observatory, France. \\
Holl,\ B.                       & Department of Astronomy, University of Geneva, Switzerland. \\
Hourihane,\ A.P.                & Institute of Astronomy, University of Cambridge, UK. \\ 
Huber,\ D.                      & University of Sydney, Australia. \\
Hunt,\ J.A.S.                   & Dunlap Institute for Astronomy and Astrophysics, University of Toronto, Canada. \\
Jasniewicz,\ G.                 & Universit\'e Montpellier, France. \\ 
Johansen,\ A.                   & Lund Observatory, Sweden. \\
Jonker,\ P.G.                   & SRON Netherlands Institute for Space Research, The Netherlands. \\
Jordan,\ S.                     & ARI/ZAH, University of Heidelberg, Germany. \\ 
Kawata,\ D.                     & Mullard Space Science Laboratory, University College London, UK. \\
Kervella,\ P.                   & Paris Observatory, France. \\
Kontizas,\ E.                   & National Observatory of Athens, Greece.\\
Kontizas,\ M.                   & National and Kapodistrian University of Athens, Greece.\\
Kordopatis,\ G.                 & Leibniz-Institut f\"ur  Astrophysik Potsdam (AIP), Germany. \\ 
Korn,\ A.J.                     & Uppsala University, Sweden. \\ 
Kos,\ J.                        & University of Sydney, Australia. \\
Kotak,\ R.                      & Queen's University Belfast, U.K. \\
Krone-Martins,\ A.              & Universidade de Lisboa, Portugal.\\
Le Poncin-Lafitte,\ C.          & SYRTE, Observatoire de Paris, France.\\
Li,\ T.                         & University of Sydney, Australia. \\
Loinard,\ L.                    & Instituto de Radioastronomía y Astrofísica, UNAM, Mexico. \\
Lucas,\ P.W.                    & University of Hertfordshire, UK. \\ 
Luri,\ X.                       & Universitat de Barcelona, ICCUB-IEEC , Spain. \\ 
Lynas-Gray,\ T.                 & University of Oxford, UK. \\
Madsen,\ M.B.                   & Niels Bohr Institute, University of Copenhagen, Denmark. \\  
Magrini,\ L.                    & INAF-Osservatorio di Arcetri. Italy. \\
Mahabal,\ A.                    & Astronomy Department, Caltech, USA.\\
Makarov,\ V.V.                  & US Naval Observatory, USA. \\
Manteiga,\ M.                   & Universidade da A Coru\~na, A Coru\~na, Spain.\\
Marconi,\ M.                    & INAF-Osservatorio Astronomico di Capodimonte, Italy. \\ 
Marshall,\ D.J.                 & CEA-Saclay / University Paris Diderot, France. \\ 
Martin-Fleitas,\ J.             & Aurora Technology B.V., Spain. \\
Martin,\ E.L.                   & Centro de Astrobiologia, Spain. \\ 
Minchev,\ I.                    & Leibniz-Institut für Astrophysik Potsdam (AIP), Germany. \\
Moitinho,\ A.                   & CENTRA - University of Lisbon, Portugal. \\
Montes,\ D.                     & Universidad Complutense de Madrid, Spain. \\ 
Muinonen,\ K.                   & University of Helsinki and National Land Survey, Finland. \\ 
Muraveva ,\ T.                  & INAF-Osservatorio Astronomico di Bologna, Italy. \\    
Musella,\ I.                    & INAF-Osservatorio Astronomico di Capodimonte, Italy. \\ 
Nascimbeni,\ V.                 & Universit\`a di Padova, Italy. \\ 
Nicastro ,\ L.                  & INAF-IASF Bologna, Italy. \\
Nienartowicz\, K.               & Geneva Observatory, Switzerland. \\ 
Nieva,\ M.F.                    & University of Innsbruck, Austria. \\
Parker,\ R.J.                   & Liverpool John Moores University, UK.\\
Pauwels,\ Th.                   & Koninklijke Sterrenwacht van België, Belgium.  \\ 
Pfenniger,\ D.                  & Geneva Observatory, Switzerland. \\
Pourbaix,\ D.                   & FNRS, Universite Libre de Bruxelles, Belgium.\\
Pr\v sa,\ A.                    & Villanova University, USA. \\
Ramos-Lerate,\ M.               & Vitrociset, Beligium. \\
Randich,\ S.                    & INAF-Osservatorio Astrofisico di Arcetri, Italy. \\ 
Read\, J.I.                     & University of Surrey, UK. \\
Reffert,\ S.                    & LSW/ZAH, University of Heidelberg, Germany. \\
Regoes,\ E.                     & Wigner Center for Physics, Budapest, Hungary \\
Reipurth,\ B.                   & University of Hawaii, USA. \\
Reyl\'e,\ C.                    & Besan\c{c}on Observatory, France.\\
Ripepi,\ V.                     & INAF-Capodimonte Observatory, Italy. \\
Riva,\ A.                       & INAF - Osservatorio Astrofisico di Torino, Italy. \\
Robin,\ A.                      & Institut UTINAM,  OSU THETA, Besan\c{c}on, France. \\ 
Rowell ,\ N.                    & University of Edinburgh, UK. \\
Ruiz-Dern,\ L.                  & Paris Observatory, France. \\
Sacco, G.G.                     & INAF-Osservatorio Astrofisico di Arcetri, Italy. \\
Sahlmann,\ J.                   & Space Telescope Science Institute, USA. \\
Santana-Ros,\ T.                & Astronomical Observatory of Adam Mickiewicz University in Poznan, Poland \\
Schoedel,\ R.                   & Instituto de Astrof\'isica de Andaluc\'ia (CSIC), Spain. \\
Schultheis,\ M.                 & Observatoire de la  Cote d'Azur, Nice, France. \\ 
Seabroke,\ G.M.	                & Mullard Space Science Laboratory, University College London, UK.\\
Sharma,\ S.                     & University of Sydney, Australia. \\
Siddiqui,\ H.                   & Telespazio Vega, UK.\\
Siebert,\ A.                    & Observatoire astronomique de Strasbourg, France. \\
Simón-Díaz,\ S.                 & Instituto de Astrofísica de Canarias, Spain. \\
Smart,\ R.L.                    & Osservatorio Astrofisico di Torino, Italy. \\ 
Smith,\ M.C.                    & Shanghai Astronomical Observatory, China. \\
Sousa,\ S.G.                    & Instituto de Astrofísica e Ciências do Espaço, Universidade do Porto, Portugal. \\
Sozzetti,\ A.                   & INAF - Osservatorio Astrofisico di Torino, Italy. \\ 
Spoto,\ F.                      & Observatoire de la Cote d'Azur, France. \\
Steele,\ I.A.                   & Liverpool John Moores University, UK.  \\
Steinmetz,\ M.                  & Leibniz Institute for Astrophysics Potsdam (AIP), Germany. \\
Stello,\ D.                     & University of Sydney, Australia. \\
Szabó,\ R.                      & Konkoly Observatory, MTA CSFK, Hungary. \\
Tautvaisiene,\ G.               & Vilnius University, Lithuania. \\
Teixeira,\ R.                   & IAG-University of S\~{a}o Paulo, Brazil.\\
Tévenin,\ F.                    & Observatoire de la Côte d'Azur, France. \\
Thuillot, \ W.                  & Paris Observatory, France. \\
Torra,\ J.                      & Institut de Ci\`encies del Cosmos, ICCUB-IEEC, Spain. \\
Trager,\ S.C.                   & Kapteyn Astronomical Institute, University of Groningen, Netherlands. \\
Vaccari,\ M.                    & University of the Western Cape, South Africa. \\
Valentini,\ M.                  & Leibniz-Institut f\"ur Astrophysik Potsdam (AIP), Germany.\\
van Altena,\ W.F.               & Yale University, USA. \\
van den Heuvel,\ E.P.J.         & University of Amsterdam, Holland. \\ 
van Velzen,\ S.                 & The Johns Hopkins University, USA. \\
Villaver,\ E.                   & Universidad Autonoma de Madrid, Spain. \\
Voss,\ H.                       & Universitat de Barcelona, Spain. \\ 
Walton,\ N.A.                   & IoA, University of Cambridge, UK. \\ 
Worley,\ C.C.                   & Institute of Astronomy, University of Cambridge, U.K. \\ 
Wright,\ N.J.                   & Keele University, UK. \\
Zwitter,\ T.                    & University of Ljubljana, Slovenia.
\end{longtable}

\section{Other supporting scientists -- added after proposal submission}\label{Sec:AppC}
\begin{longtable}[l]{ll}
	Alfaro,\ E.J.               & Instituto de Astrofísica de Andalucía, Granada, Spain. \\
	Bailer-Jones,\ C.           & Max Planck Institute for Astronomy, Heidelberg, Germany. \\
	Berihuete,\ A.              & Universidad de C\'adiz, Spain. \\
	Lebreton,\ Y.               & Paris-Meudon Observatory, France. \\
	Liu,\ C.                    & National Astronomical Observatories, China. \\ 
	Messina,\ S.                & INAF-Catania Astrophysical Observatory, Italy. \\
	Molinaro,\ R.               & INAF-Osservatorio Astronomico di Capodimonte, Napoli, Italy. \\
	Moniez,\ M.                 & Laboratoire de l’Accélérateur Linéaire, Université Paris-Sud, Université Paris-Saclay, France. \\
	Pagano,\ I.                 & INAF-OACT, Italy. \\
	Süveges,\ M.                & University of Geneva, Switzerland/Max-Planck-Institute for Astronomy, Heidelberg, Germany. \\
	Szabados,\ L.               & Konkoly Observatory, Budapest, Hungary. \\
	Wilkinson,\ M.I.            & University of Leicester, United Kingdom. \\
\end{longtable}	

\end{document}